%% file: main.tex
\documentclass[pre,reprint,superscriptaddress,tightenlines,twocolumn,showpacs,floatfix,longbibliography,nofootinbib]{revtex4-2}
\pdfoutput=1

\input{packages}

\begin{document}
\title{A Minimal Framework for Optimizing
Vaccination Protocols \\ Targeting Highly Mutable Pathogens}

\author{Saeed Mahdisoltani}
\email{saeedmah@mit.edu}
\affiliation{Department of Chemical Engineering, Massachusetts Institute of Technology, Cambridge, MA 02139}
\affiliation{Institute for Medical Engineering and Science, Massachusetts Institute of Technology, Cambridge, MA 02139}
\affiliation{Department of Physics, Massachusetts Institute of Technology, Cambridge, MA 02139}
\author{Pranav Murugan}
\affiliation{Department of Physics, Massachusetts Institute of Technology, Cambridge, MA 02139}
\affiliation{Department of Electrical Engineering and Computer Science, Massachusetts Institute of Technology, Cambridge, MA 02139}
\author{Arup K. Chakraborty}
\email{arupc@mit.edu}
\affiliation{Department of Chemical Engineering, Massachusetts Institute of Technology, Cambridge, MA 02139}
\affiliation{Institute for Medical Engineering and Science, Massachusetts Institute of Technology, Cambridge, MA 02139}
\affiliation{Department of Physics, Massachusetts Institute of Technology, Cambridge, MA 02139}
\affiliation{Ragon Institute of Massachusetts General Hospital, Massachusetts Institute of Technology and Harvard University, Cambridge, MA 02139}
\affiliation{Department of Chemistry, Massachusetts Institute of Technology, Cambridge, MA 02139}
\author{Mehran Kardar}
\email{kardar@mit.edu}
\affiliation{Department of Physics, Massachusetts Institute of Technology, Cambridge, MA 02139}


\begin{abstract}
    \input{0_abstract.tex}

\end{abstract}

\maketitle

\section{Introduction}	\label{sec:introduction}

\input{1_introduction.tex} 

\section{Theoretical model} \label{sec:model}
    \input{2A_model.tex}

    \input{2B_mean-field.tex}


\section{Analytical results}  
\label{sec:analytical}
    \input{3A_path-integral}

    \input{3B_operator-approximation}

\section{Computational results}
\label{sec:computational}
    \input{4_computational}

\section{Discussion}    \label{sec:discussion}
    \input{5_discussion}

\section{Acknowledgments}
We would like to thank Leerang Yang and Federica Ferretti for helpful discussions. This research was partially supported by NIH grant 5R01AI175489-02 and the Ragon Institute of MGH, MIT, and Harvard.
MK also acknowledges support from NSF grant DMR-2218849.

\textit{Author contributions --}
S.M., A.K.C., and M.K. designed the research. S.M. carried out the analytical
calculations, numerical analysis, and stochastic simulations. P.M. provided the original code for Gillespie simulations. S.M., M.K., and A.K.C. analyzed and discussed the results and wrote the paper. PM offered comments on the manuscript.

\textit{Declaration of interests --}
A.K.C. is a consultant (titled Academic Partner) for Flagship Pioneering and
also serves on the Strategic Oversight Board of its affiliated company, Apriori
Bio, and is a consultant and SAB member of another affiliated company, Metaphore Bio.

\appendix

\section{Derivation of the mean-field in the continuum limit} \label{appendix:master-eq}
    \input{Appendix_master}

\section{Details of the path integral calculation}    \label{appendix:path-integral}
    \input{Appendix_path-integral}

\section{BCH approximation for dynamics on discrete bins} \label{appendix:bch-discrete}
    \input{Appendix_BCH-discrete}

\input{Appendix_trajectory}

\clearpage
\bibliography{BCH-PRE} 

\end{document}

%% file: packages.tex
\usepackage{latexsym,amsmath,amsfonts,amssymb,bm,empheq} 
\usepackage{physics}
\usepackage{nicefrac} 
\usepackage[dvipsnames]{xcolor}
\usepackage{booktabs}
\usepackage{multirow}
\usepackage{tikz}   
\usetikzlibrary{arrows.meta,decorations.markings,bending}
\usepackage{pgfplots}
\usepackage{subfigure}

\definecolor{nblue}{RGB}{28,130,185}
\definecolor{cgreen}{RGB}{76,153,0}
\definecolor{myorange}{RGB}{245,156,74}

\usepackage{hyperref}
\hypersetup{
  colorlinks=true,
  citecolor=magenta,
  urlcolor=-myorange}

\usepackage{array}
\AtBeginDocument{
\cmidrulewidth=.03em
}

%% file: 0_abstract.tex
A persistent public health challenge is finding immunization schemes that are effective in combating highly mutable pathogens such as HIV and influenza viruses. To address this, we analyze a simplified model of affinity maturation, the Darwinian evolutionary process B cells undergo during immunization. The vaccination protocol dictates selection forces that steer affinity maturation to generate antibodies. We focus on determining the optimal selection forces exerted by a generic time-dependent vaccination protocol to maximize production of broadly neutralizing antibodies (bnAbs) that can protect against a broad spectrum of pathogen strains. The model lends itself to a path integral representation and operator approximations within a mean-field limit, providing guiding principles for optimizing time-dependent vaccine-induced selection forces to enhance bnAb generation. We compare our analytical mean-field results with the outcomes of stochastic simulations and discuss their similarities and differences.

%% file: 1_introduction.tex
The adaptive immune system has a remarkable ability to detect and combat a virtually unlimited number of previously unseen pathogens~\cite{victora-review-2022,walczak-review-2020}. Immune responses are mediated by T lymphocytes (T cells) and B lymphocytes (B cells), each equipped with surface receptors known as T cell receptors (TCRs) and B cell receptors (BCRs), respectively. 
In a healthy human adult, there are approximately $10^{11}$ T cells and B cells distributed throughout the body~\cite{sender2023total}. 
With the staggering number of TCR and BCR sequences ($> 10^{14}$) generated via the V(D)J recombination process~\cite{yaari2015practical}, most T cells and B cells express a unique surface receptor distinct from others. This extensive receptor diversity enables the immune system to mount specific responses against each new pathogen encountered, including those that emerge after an individual's birth.

Upon exposure to a new pathogen or vaccine component (collectively referred to as antigens), B cells and antibodies with high binding affinity for the surface proteins of the pathogen are generated by a process called \textit{affinity maturation}~\cite{victora-GC-2016,victora-review-2022}. 
Affinity maturation entails an accelerated Darwinian evolution of naive B cells, and it occurs within specialized structures called germinal centers (GCs), which are transiently formed in secondary lymphoid organs such as lymph nodes~\cite{victora-review-2012}. 
First, naive B cells undergo activation and become a candidate for entry into GCs if their BCRs can bind sufficiently strongly to specific residues known as \textit{epitopes} on the surface proteins of the antigen~\cite{victora-GC-2016}. 
Upon entering the GC, the activated B cells undergo rapid replication while concurrently accumulating somatic mutations into their BCRs at an accelerated rate. 
The GC B cells then interact with the infecting antigen, which is displayed on Follicular Dendritic Cells (FDCs) located within the GC environment.  
B cells whose receptors exhibit stronger binding to the displayed antigen have a greater probability of internalizing the antigen and are consequently more likely to undergo positive selection. Positive selection involves multiple stages of productive interactions of the B cells with certain T cell types within the GC. 
Conversely, B cells with low-affinity BCRs are typically eliminated via apoptosis as they are less likely to successfully compete with high-affinity B cells for the limited selection signals~\cite{basic-immunology}. 
The majority of positively selected GC B cells undergo multiple rounds of replication, somatic mutation, and selection processes. As a result of the competition among GC B cells during the affinity maturation process, an initially naive B cell population with low-affinity BCRs gradually evolves to exhibit strong binding to the antigen. 
During each round, a small number of the positively selected B cells exit the GCs and differentiate into either memory B cells or plasma cells~\cite{victora-GC-2016,basic-immunology}. 
Plasma cells are responsible for secreting antibodies, which are soluble forms of the BCR that can bind the antigen's surface proteins with high affinity and neutralize the pathogen's ability to infect host cells. Thus, the affinity maturation process generates an effective immune response by producing antibodies that specifically neutralize a particular pathogen~\cite{victora-GC-2016}.

Memory B cells, on the other hand, play a crucial role in the immune system by retaining a memory of the antigen that activated their parent B cell during the initial exposure. This memory enables a rapid immune response upon re-encountering the same antigen in the future~\cite{basic-immunology}. 
Upon re-exposure to a pathogen, existing memory B cells are selected in an affinity-dependent manner and rapidly expanded outside GCs, a process akin to the mechanisms occurring within GCs but with little to no mutation~\cite{victora-review-2022}. A significant fraction of these expanded memory B cells differentiate into plasma cells,  secreting a surge of antibodies that offer immediate protection. Meanwhile, new GCs are also formed, producing even higher affinity memory B cells and antibodies over more extended time scales, thus bolstering the immune response against the recurring pathogen.

Vaccines leverage immunological memory to confer protection against severe or life-threatening illnesses caused by specific pathogens they target~\cite{vax-fundamentals,vax-review-2003}.
By eliciting memory cells and antibodies, vaccines prime the immune system to mount a rapid and specific response to the targeted pathogen upon subsequent exposures~\cite{vax-fundamentals}. 
While vaccines have significantly improved in terms of safety and efficacy over time, developing effective vaccines against highly mutable pathogens remains a continuing challenge. Pathogens such as HIV, influenza, and SARS-CoV-2 mutate rapidly, allowing them to evade the immune responses targeted at specific strains.  
Surface proteins of highly mutable viruses can exhibit significant variation across different strains, rendering vaccine-induced antibodies that neutralize a specific strain ineffective against other strains. This variability complicates vaccine development efforts, as vaccines must contend with the ever-evolving nature of these pathogens.

Certain residues within the surface proteins of highly mutable viruses must remain relatively conserved across mutant strains to preserve the virus's ability to infect host cells. 
Antibodies targeting these conserved residues hold the potential to neutralize a broad spectrum of virus strains~\cite{burton-review-2016}. 
In some HIV-infected patients, broadly neutralizing antibodies (bnAbs) naturally evolve, demonstrating the immune system's capability to elicit such cross-reactive antibodies~\cite{bnab-prevalence-2014,burton-review-2016}. 
However, bnAbs are seldom generated in response to natural infection or conventional vaccination, and their emergence typically occurs over extended periods and in limited quantities~\cite{bnab-prevalence-2014,laursen2013broadly,kreer-bnab_prob-2023}. 
Vaccination strategies aimed at eliciting bnAbs have therefore emerged as a promising approach for diseases like HIV and influenza~\cite{burton-review-2016, burton-bnab-2018, laursen2013broadly}. 
These vaccines have the potential to confer protection against variant strains by inducing a robust immune response targeted at conserved epitopes. 
Despite the advancements, designing universal variant-proof vaccines for mutating pathogens remains a formidable challenge, as the typical immune response to such pathogens is predominantly driven by strain-specific B cells and antibodies.

For modeling purposes, we can conceptualize the affinity maturation process as the evolutionary population dynamics of B cells `species' within GCs, characterized by inherent stochasticity stemming from randomness in various steps including B cell activation, GC entry, replication, antigen internalization, T cell selection, and more.  
In this framework, vaccine antigens serve as external interventions  guiding this stochastic dynamics process towards desired goals. Designing vaccination protocols that confer broad protection against mutant viruses entails selecting appropriate external factors that steer GC processes to increase the likelihood of rare evolutionary trajectories leading to the development of bnAbs. For instance, external manipulations can involve immunization with variant antigens sharing conserved regions of a virus' spike while differing in variable parts. Experimental and computational investigations have focused on enhancing bnAb production through carefully designed vaccination protocols~\cite{burton-review-2016, barouch-2020,amitai-population-2017,kayla-bnab-2020,ganti-bnab-2021}, but achieving success remains elusive. Albeit constrained, there are various routes to guiding affinity maturation, including use of variant antigens, number of vaccine doses, timing between shots, antigen composition in each shot, and dosage. The large space of variables for choosing vaccination protocols makes an unguided search unlikely to succeed. 

Computational approaches have proven invaluable in systematically exploring the large space of vaccination protocols, providing insights into the outcomes of affinity maturation in response to diverse vaccine designs~\cite{akc-perspective-2017}. 
Several computational studies have investigated the efficacy of sequential and cocktail immunization procedures for eliciting bnAbs, highlighting an optimal difference between  the variable parts of antigen variants included in vaccine shots that maximize the likelihood of bnAb production~\cite{wang-bnab-2015,shaffer2016optimal,kayla-bnab-2020,ganti-bnab-2021}. 
Signatures of such optimal antigen differences have been observed in HIV patients who naturally develop bnAbs~\cite{burton-plos,jardine2015priming} and in mice vaccinated by sequentially administered influenza antigens~\cite{amitai2020defining,dong2018cross}. 
Mechanistic explanations for why an optimal difference between variant antigens promotes bnAb evolution have also been proposed~\cite{shaffer2016optimal,wang-bnab-2015,kayla-bnab-2020,ganti-bnab-2021}. 
However, existing studies typically focus on immunization with a fixed  cocktail of antigens~\cite{wang-bnab-2015,shaffer2016optimal} or  sequential procedures where the variable parts of the antigens are substantially different, maximizing the selection pressure in favor of bnAbs~\cite{kayla-bnab-2020,ganti-bnab-2021}. 

In this paper, we use analytical and computational methods to model the complex task of designing vaccines capable of eliciting immune responses that confer protection against diverse mutant strains within a virus family. 
To unveil guiding principles for optimizing bnAb production, we investigate generic time-dependent vaccination schemes that involve administering mixtures of relatively similar antigens over time. 
Our model significantly expands upon previous works in Refs.~\cite{ganti-bnab-2021}~and~\cite{kayla-bnab-2020} by allowing variations in both the location and spread of the fitness landscape induced by the vaccination protocol. To describe BCR-antigen interactions, we adopt a shape-space representation~\cite{perelson-clonal-1979, perelson-97-deriving}, which enables us to capture the essence of the evolutionary dynamics of B cell populations in response to vaccination in a minimal framework. 

The rest of the paper is organized as follows: In Section~\ref{sec:model}, we describe a simplified model that captures the essential steps of affinity maturation of the B cell population in GCs, namely replication, apoptosis, and mutations. After developing a mean-field approximation to the stochastic dynamics, Sections~\ref{sec:path-integral}~and~\ref{sec:bch} provide optimization results based on a path integral representation, and operator approximation methods. In particular, for a Gaussian initial germline B cell population that is away from the bnAb state, we derive an exact identity (Eq.~\eqref{eq:nfx0-ratio-formula}) that directly provides the mean-field prediction for the optimal location for the focus of the fitness landscape imposed by the vaccine (Eq.~\eqref{eq:xstar-def}). Using this identity, along with operator methods for approximating the solution to the population dynamics, we obtain a result (Eq.~\eqref{eq:nfx-BCH-simplified}) that enables us to investigate the optimal spread of the vaccine fitness. We next 
compare the mean-field predictions with stochastic simulations in Section~\ref{sec:computational}. Finally, we present concluding remarks and outline future research directions in Section~\ref{sec:discussion}. 
There are four appendices that contain further information regarding the derivation of the continuum mean-field model (Appendix~\ref{appendix:master-eq}), the details of the path integral calculation (Appendix~\ref{appendix:path-integral}), the operator approximation for the discrete-bin dynamics (Appendix~\ref{appendix:bch-discrete}), and different aspect of the B cell dynamics under different vaccine protocols (Appendix~\ref{appendix:trajectory}). 

%% file: 2A_model.tex
\begin{figure}[t]
    \centering
    \includegraphics[width=\linewidth]{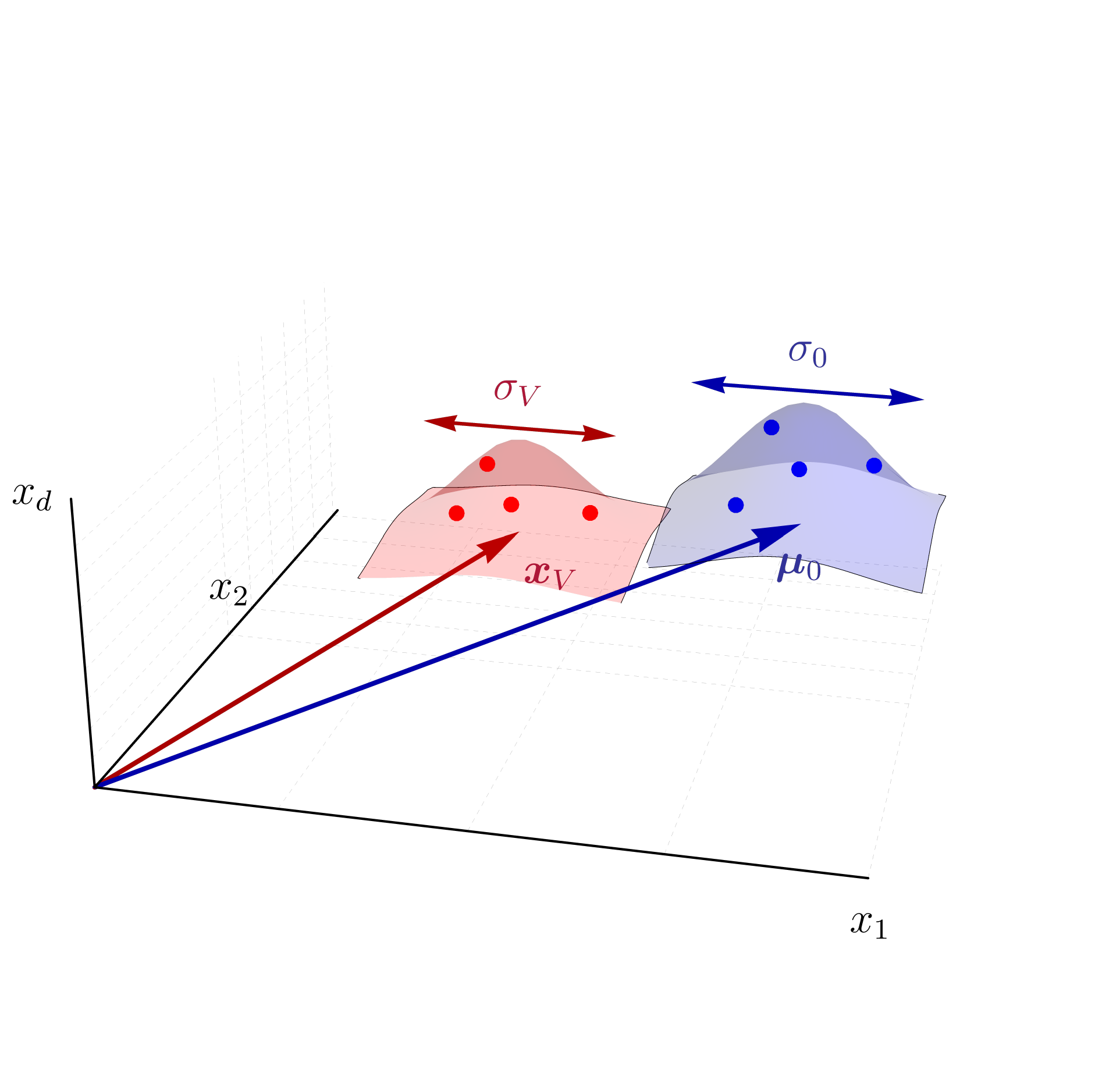}
    \caption{Schematic depiction of a $d$ dimensional shape space. The germline B cell population $n_0(\bm{x})$ (in blue) is centered at $\bm{\mu}_0$ with a spread of $\sigma_0$, while the vaccine-induced fitness profile (in red) is parameterized by the its center $\bm{x}_V$ and spread $\sigma_V$. 
    The blue and red dots denote specific antibodies and antigens, respectively, and the bell-like surfaces are the coarse-grained approximations of the corresponding density profiles.  
    Given a constrained total vaccine dose (Eq.~\eqref{eq:constraint}), the objective is to determine a time-dependent vaccination profile that optimally shifts the B cell population in  shape space, ultimately maximizing the final bnAb count corresponding to BCRs located at $\bm{x}=0$. } 
    \label{fig:schematic}
\end{figure}
\subsection{Minimal model for BCR evolution} 
\label{sec:minimal-model}

To construct a mechanistic model of affinity maturation, we frame it as the population dynamics governing the evolution of B cells--or equivalently, their BCRs--where replication rates are determined by their binding affinity to the stimulating antigen. 
The interaction between BCRs and antigens is typically captured through phenomenological approaches such as the string model~\cite{wang-bnab-2015,kayla-bnab-2020,perelson-2015}. 
Here we adopt a more coarse-grained approached based on the concept of \textit{shape-space} representation~\cite{perelson-clonal-1979,perelson-97-deriving,shaffer2016optimal}.  
In particular, BCRs and antigens are represented by points in a  $d$-dimensional Euclidean space (Fig.~\ref{fig:schematic}). Each dimension corresponds to parameters pertinent to calculating BCR-antigen interactions, including factors such as amino acids sequences, spatial conformations, hydrophobicity, and the like. 
As these coordinates in the shape space can assume distinct values, we initially discretize the shape space into a number of \textit{similarity} bins indexed by $\bm{x}=(x_1,x_2,\ldots,x_d)$. 

BCRs and antigens positioned closely in shape space are presumed to have complementary characteristics that allow them to bind strongly, while those situated farther apart exhibit weaker binding~\cite{perelson-clonal-1979,perelson-review-1997}. 
Following the approach outlined in Ref.~\cite{shaffer2016optimal}, we characterize the binding free energy between a BCR and an antigen located in similarity bins $\bm{R}_{\mathrm{BCR}}$ and $\bm{R}_{\mathrm{Ag}}$ by their distance as  
$$E_{\mathrm{bind}} \propto k_{\mathrm{B}}T \,  ||\bm{R}_{\mathrm{BCR}} -\bm{R}_{\mathrm{Ag}}||^2.$$ 
Consequently, the associated binding affinity is expressed as $(E_{\mathrm{thr}}-E_{\mathrm{bind}})$, with $E_{\mathrm{thr}}$ denoting the threshold binding free energy for B cell activation. 
B cells bind to the antigen presented on the FDCs with an affinity determined by the equilibrium constant of the BCR-antigen binding, defined as  
$K_a = \exp\left[  (E_{\mathrm{thr}}-E_{\mathrm{bind}})/{k_{\mathrm{B}}T} \right]$. 
Increased affinities correspond to  stronger BCR-antigen binding with higher equilibrium constants. Affinity maturation can amplify the binding affinity of naive BCRs by tenfold and its associated $K_a$ by a thousandfold~\cite{yang1999mutational}. 
Furthermore, we assume that BCRs located closer to the origin ($\bm{x}=0$) in the shape space have greater affinity toward the conserved residues of the antigen surface proteins. 
These BCRs exhibit increased  tolerance to mutations in the surrounding variable residues of the antigen, facilitating the elicitation of bnAbs. Our objective, therefore, is to devise a vaccination procedure conducive to directing affinity maturation to optimally guide the initially naive BCRs toward the $\bm{x}=0$ bin.

Since high-affinity B cells are more likely to undergo positive selection and further replication, the binding affinity to the vaccine antigen(s) effectively imposes a \textit{fitness function} on GC B cells. 
When a single antigen type is included in a vaccine shot, the induced fitness function exhibits a sharply peaked distribution centered around the complementary BCR bin within the shape space. This distribution features a minimal spread, denoted as $\sigma_{\min}$, to encompass nearby bins whose BCRs may also bind the given antigen, albeit with reduced affinity. 
Conversely, in the case of a cocktail shot containing mixtures of different antigen types, the overall fitness function results from the summation of the individual fitness functions induced by each antigen type.\footnote{We are assuming a `see all' scenario wherein all antigens are well-mixed and presented homogeneously on FDCs, such that B cells encounter all the vaccine antigens at every round~\cite{shaffer2016optimal}.} 
This typically yields a non-convex fitness landscape with multiple peaks and valleys. 
However, if the antigens included in a cocktail shot are sufficiently similar (e.g.,  sharing a common epitope containing conserved residues and surrounding variable portions that are different), the individual fitness peaks would cluster closely in shape space. 
Henceforth, we assume the administered antigens at each time point are sufficiently similar to allow for approximating the overall vaccine-induce fitness landscape with a smooth convex function encompassing these peaks (Fig.~\ref{fig:schematic}). The resultant coarse-grained fitness function is then roughly centered in the midst of the individual antigens, with a spread generally narrower for more similar antigens. 
We represent this coarse-grained fitness function as $V(\bm{x})$, with its peak located at $\bm{x}_V$ (referred to as the vaccine center or focus) and its spread denoted as $\sigma_V$. 
It is worth noting that under the assumption of similar antigens, $\bm{x}_V$ may generally be distant from the BCR bin associated with the bnAb sequences. 
%

Now let us describe a minimal model for the affinity maturation process within the representation outlined above. 
The germline distribution that initially seeds the GCs is denoted by $n(\bm{x},t=0) \equiv n_0(\bm{x})$. As affinity maturation progresses, this distribution evolves over time to $n(\bm{x},t)$. We model the evolution of BCRs during affinity maturation through the following stochastic processes: 
\begin{itemize}
    \item \textit{Replication:}  the rate of B cell replication is determined by the stimulating antigen(s) included in the vaccine dose. By altering the composition of the presented antigens, we assume it is possible to modulate the vaccine-induced fitness, $V(\bm{x},t)$, across different bins in  shape space, with corresponding time-dependent vaccine center and spread,  $\bm{x}_V(t)$ and $\sigma_V(t)$.  
    Note that we introduce a time dependence in $V$ to accommodate vaccination protocols where different antigens or cocktails of antigens are introduced at different times.  
    \item \textit{Mutation:} BCR mutations occur with rate $\gamma_{\bm{x},\bm{y}}$, corresponding to jumps between shape-space bins $\bm{x}$ and $\bm{y}$. 
    While nucleotide mutations can theoretically move a BCR between any two similarity bins, the coarse-grained nature of shape-space coordinates--reflecting the effect of many amino acids on the binding affinity--suggests that large jumps are exceedingly rare.
    For the simplified mean-field model in the next section, we will assume mutational jumps are local and only occur between nearby bins. This assumption allows us to model them by the diffusion of BCRs in the appropriate continuum limit.   
    \item  \textit{Apoptosis:}  B cells die with rate $\lambda_{\bm{x}}$, representing their baseline apoptotic tendency within GCs when not receiving positive selection signals~\cite{victora-GC-2016,liu1989mechanism}. 
\end{itemize}  

The rules outlined above describe a generalized linear birth-death-mutation process and are encapsulated in the following Fock space master equation~\cite{grassberger1980fock,tauber}
\begin{align} \label{eq:master-fock}
    \partial_t P(\lbrace n \rbrace , t) = 
    \quad &\sum_{\bm{x}} V_{\bm{x}}(t) \, \big[ \mathbb{E}_{\bm{x}}^{-}-1 \big] \big(n_{\bm{x}} P(\lbrace n \rbrace ,t) \big)
        \nonumber\\
    \quad+&\sum_{\bm{x}, \bm{y}} \gamma_{\bm{x},\bm{y}} 
    \big[ \mathbb{E}^+_{\bm{x}} \mathbb{E}_{\bm{y}}^{-} -1 \big] 
    \big( n_{\bm{x}} \, P(\lbrace n \rbrace , t) \big) 
    \, 
    \nonumber\\
    \quad+&\sum_{\bm{x}} \lambda_{\bm{x}} \,
    \big[\mathbb{E}^+_{\bm{x}} - 1\big] \big(n_{\bm{x}}  P(\lbrace n \rbrace ,t) \big) \, .
\end{align}
Here, the BCR population is denoted by
$\lbrace n \rbrace = \lbrace n_1, n_2 \, \ldots \rbrace$ 
where $n_i$ is a non-negative integer representing the occupation number of the $i$th bin, and $P(\lbrace n \rbrace, t)$ is the probability of having that specific population at time $t$. 
On the right-hand side of Eq.~\eqref{eq:master-fock}, we introduce the operators $\mathbb{E}_{\bm{x}}^\pm$, defined as  
$$\mathbb{E}_{\bm{x}}^{\pm} \big(n_{\bm{y}} P(\lbrace n \rbrace , t) \big) \equiv (n_{\bm{y}} \pm \delta_{\bm{x},\bm{y}}) \, P(\lbrace n\rbrace \pm \mathbb{I}_{\bm{x}}) \, ,$$ 
where $\bm{x}$ and $\bm{y}$ are bin indexes, $\delta_{\bm{x},\bm{y}}$ is the Kronecker delta,  and $\lbrace n \rbrace \pm \mathbb{I}_{\bm{x}}$ denotes a population that differs from $\lbrace n \rbrace$ by one more or less B cell residing in bin $\bm{x}$.  

Each line on the right-hand side of Eq.~\eqref{eq:master-fock} represents the total rate of change of population probabilities (i.e., gain minus loss) caused by replication, mutation, and apoptosis events, respectively.  
It is worth noting that by describing the occupancy numbers of the shape-space bins, the solution to Eq.~\eqref{eq:master-fock} would provide a probabilistic description of the entire BCR population in the high-dimensional shape space.

%% file: 2B_mean-field.tex
\subsection{Mean-field dynamics and the continuum limit}    
\label{sec:mean-field-model}

The abstract model described above offers a simplified perspective on affinity maturation. However, deriving insights from the resulting master equation (Eq.~\eqref{eq:master-fock}) is challenging due to the exponentially large Fock space and the interconnectedness of the time-evolution for all configurations $\{n\}$. 
To address this, and obtain an analytically tractable formulation, we consider the dynamics of population averages, denoted by
$\expval{n_{\bm{x}}(t)} = \sum_{\lbrace n \rbrace} n_{\bm{x}} P(\lbrace n \rbrace,t)$, and, subsequently, taking the continuum limit of the shape space. 
This leads to the following continuum mean-field equation (see Appendix~\ref{appendix:master-eq} for derivation) 
\begin{align}       \label{eq:mean-field}
    \partial_t n(\bm{x},t) = 
    D \, \nabla^2 n(\bm{x},t)
    + \left[V(\bm{x},t)- \lambda \right] \, n(\bm{x},t)  \, ,
\end{align}
where, for notational simplicity, we retain the symbol $n$ to represent the average population \textit{density} in the continuum limit of $\bm{x}$.  
We also assume the death rate $\lambda$ and the diffusion coefficient $D$ (obtained from the underlying mutation rates via Eq.~\eqref{eq:mu-to-D}) are constants independent of $\bm{x}$.  
Note that, as described earlier,   we consider vaccines with either single antigen types or cocktails wherein the administered antigens are not substantially different from each other. Consequently, at any given time point, the most cross-reactive BCRs to the administered vaccine antigens--namely, BCRs located at $\bm{x}_V(t)$ in the center of the fitness--may not correspond to a bnAb sequence. This is somewhat distinct from the setting described in Ref.~\cite{ganti-bnab-2021}, where antigens are different enough that bnAb sequences are always the most fit. 

Solving the mean-field dynamics (Eq.~\eqref{eq:mean-field}) requires specifying the initial germline distribution, which typically has little overlap with the target bin in  shape space. 
We thus assume $n_0(\bm{x})$ is centered at a distant location $\bm{\mu}_0$ and has a width $\sigma_0$ that is an indication of the diversity of the germline BCRs (see Fig.~\ref{fig:schematic}). 
In the spirit of the central limit theorem, we model $n_0(\bm{x})$ by a normal distribution
\begin{align}   \label{eq:n0-def}
    n_0(\bm{x}) = N_0 \, 
    \dfrac{e^{-\frac{(\bm{x}-\bm{\mu}_0)^2}{2\sigma_0^2}}}{\left(2\pi \sigma_0^2\right)^{d/2}} \equiv N_0 
    \mathcal{G}(\bm{x}-\bm{\mu}_0, \sigma_0^2)\, ,
\end{align}
where  $\mathcal{G}$ denotes the Gaussian function. 
Here, $N_0$ represents the total number of activated B cells at $t=0$.   

Our objective is to determine an optimal vaccination strategy, characterized by  
$\{ \bm{x}_V(t),\sigma_V(t)\}$,  
that maximizes the final bnAb population denoted by $n(\bm{x}=0,t_f) \equiv n_f(0)$, while adhering to a constrained total vaccine dose, i.e.
\begin{equation}    \label{eq:constraint}
    \int_0^{t_f} dt \int d^d\bm{x} \, V(\bm{x},t) = \text{constant}\,.
\end{equation}
Equation~\eqref{eq:mean-field} resembles the imaginary-time Schrodinger equation with a time-dependent potential~\cite{feynman-hibbs}. Typically, such equations lack analytical solutions, rendering a direct approach to the maximization problem infeasible.  


Before delving into the analsis of the model, we summarize the assumptions made to simplify the model and acknowledge their associated limitations:  

\begin{enumerate}
    \item We interpret Eq.~\eqref{eq:constraint} as imposing a constraint on the total amount of administered antigens in vaccine doses. 
    This interpretation implies a linear relation between the fitness function $V$ (appearing on the left-hand side of Eq.~\eqref{eq:constraint}) and the amount of the administered antigen. However, $V(\bm{x},t)$ can generally be a complicated function of the antigen concentration, incorporating nonlinear terms that arise from various factors such as competition between B cells for GC entry, positive selection within GCs, and epitope masking (see, e.g., Refs.~\cite{victora-GC-2016,batista-2022,leerang-omicron-2023,nussenzweig-2023-epitopemasking}). 
    Nonetheless, the linear relation between fitness and vaccine dose remains a plausible approximation if selection based on the amount of internalized antigen is not overly stringent. This approximation is consistent with the linear structure of the birth-death-mutation model of Eq.~\eqref{eq:master-fock}. 
    \item We assume that mutational jumps occur independently of replication events, exert incremental effects on BCR affinity, and they are equally probable in any direction (i.e., they are isotropic).    
    These assumptions underlie the symmetric form of the diffusion term in the mean-field dynamics, as included in Eq.~\eqref{eq:mean-field}, along with the vanishing drift term (see Appendix~\ref{appendix:master-eq}). 
    In a more realistic model, mutations would primarily occur during replication events and tend to be deleterious than beneficial, resulting in a coarse-grained dynamics with a nonhomogenous diffusion term and a non-vanishing drift term. In addition, deleterious mutations can also significantly reduce the affinity~\cite{leerang-omicron-2023}, potentially leading to long-ranged steps in the coarse-grained mean-field dynamics.  
\end{enumerate}

%% file: 3A_path-integral.tex
\begin{figure*}
    \centering
    \includegraphics{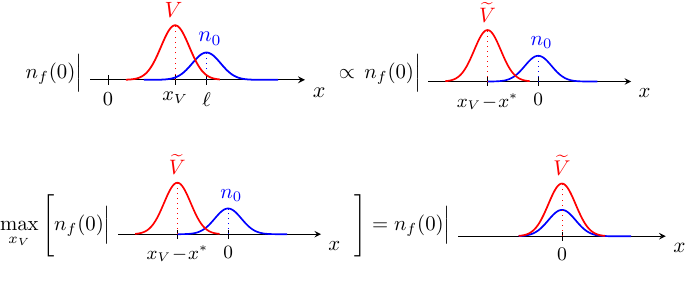}
    \caption{A visual representation of the path integral result. Top panel: the final number of bnAbs, $n_f(0)$, for a Gaussian germline $n_0$ that starts away from the target bin ($x=0$) and evolves under a general vaccine fitness $V$, is directly proportional to the final bnAbs count when $n_0$ starts at the target bin and evolves under the modified fitness $\widetilde{V}$ defined in Eq.~\eqref{eq:V-tilde-def}. The proportionality relationship is governed by Eq.~\eqref{eq:nfx0-ratio-formula}. Bottom panel: To optimize $n_f(0)$ in the mapped problem, the modified fitness $\widetilde{V}$ needs to be centered at the target, which is achieved by adjusting the optimal vaccine center according to Eq.~\eqref{eq:xv-optimal-xstar}. }
    \label{fig:PI-graphical}
\end{figure*}

As mentioned earlier, solving the mean-field equation~\eqref{eq:mean-field} explicitly for a general vaccine fitness function, and thus direct optimization of bnAbs, is not feasible. In this section, we propose a method to simplify this maximization by applying a coordinate change to the path integral representation of the solution to the mean-field dynamics (Eqs.~\eqref{eq:path-integral-solution}~and~\eqref{eq:nfx0-ratio-formula} below).

\subsection{Path integral representation and the optimal location of the fitness peak}    \label{sec:path-integral}

We begin  constructing the path integral formulation by rearranging Eq.~\eqref{eq:mean-field} for an infinitesimal timestep $\delta t$: 
\begin{align}   \label{eq:generator}
    n(\bm{x},t+\delta t) &= \{ 1 + \delta t \, D \nabla^2 + \delta t[V(\bm{x},t)-\lambda] \} n(\bm{x},t) \nonumber\\
    &=  e^{\delta t D\nabla^2} e^{\delta t[V(\bm{x},t)-\lambda]} n(\bm{x},t) + O\left(\delta t^2\right) \, .
\end{align} 
Equation~\eqref{eq:generator} provides a formal method for evolving the population profile in infinitesimal timesteps. 
By repeatedly applying Eq.~\eqref{eq:generator} to evolve the population over time, the final population $n(\bm{x},t_f) \equiv n_f(\bm{x})$ can be expressed by the formal time-ordered product~\cite{feynman-hibbs}
\begin{align}   \label{eq:trotter}
    n_f (\bm{x})= \lim_{M \to \infty}
    \prod_{j=0}^{j=M} 
     e^{\delta t D\nabla^2} \, e^{\delta t [V(\bm{x},j\delta t)-\lambda]} \, n_0(\bm{x})\,. 
\end{align}
Here, we have defined $\delta t = \frac{t_f}{M}$ with $M$ as the number of time increments, and $V_j(\bm{x})$ denotes the vaccine-induced fitness function at time $j\delta t$. 
The products in Eq.~\eqref{eq:trotter} act in the operator sense:\footnote{This product formula is also known as the first-order Suzuki--Trotter decomposition~\cite{sadrihassani,suzuki1977} and it forms the basis for the path integral formulation of quantum mechanics~\cite{feynman-hibbs}, quantum simulators~\cite{lloyd1996universal,trotter-prx}, and the so-called operator splitting methods~\cite{operator-splitting}.} 
starting from the germline population $n_0$, at the $j$th infinitesimal timestep $\delta t$, the B cell population is locally expanded/contracted according to the vaccine operator $e^{\delta t [V(\bm{x},j\delta t) - \lambda]}$, 
followed by diffusion broadening under the operator 
$e^{\delta t D \nabla^2}$.  

The effect of the diffusion operator on a test function $\eta(\bm{x})$ is given by the convolution 
$$e^{\delta t D \nabla^2} \eta(\bm{x}) = \int d^d\bm{x}' \mathcal{G}(\bm{x}-\bm{x}', 2D\delta t) \, \eta(\bm{x}'). $$
Due to the non-local nature of this term, the final population at any location in  shape space results from all potential ways that different parts of the initial population spread and expand until they reach that location at the final time. 
Upon inserting the diffusion operators in Eq.~\eqref{eq:trotter} and integrating over all intermediate points, the final population can be more explicitly expressed as 
   $$     
    \int \prod_{j=0}^{M} d^d\bm{x}_j \,
    \mathcal{G}(\bm{x}_{j+1}-\bm{x}_j, 2D\delta t) \, 
    e^{\delta t (V(\bm{x}_j, j\delta t)-\lambda)} \,
    n_0(\bm{x}_0)\,,
    $$
with $\bm{x}_{M+1} = \bm{x}$.  
This expression is then rearranged into the standard path integral form~\cite{feynman-hibbs,kardar-fields} 
\begin{align}   \label{eq:path-integral-solution}
    n_f(\bm{x}) = e^{-\lambda t_f} \int d^d\bm{x}_0 \, n_0(\bm{x}_0) 
    \int_{\bm{r}(0)=\bm{x}_0}^{\bm{r}(t_f)=\bm{x}}
    \mathcal{D} \bm{r}(u) \, 
    e^{-S[\bm{r},\dot{\bm{r}}]}\,,
\end{align}
where 
$\mathcal{D}\bm{r} \propto d\bm{x}_1 \ldots d\bm{x}_M$ stands for the measure of all paths connecting the endpoints, each weighted according to its `action' 
\begin{align}   \label{eq:path-integral-action}
S[\bm{r},\dot{\bm{r}}] = \int_0^{t_f} du \, \left( \frac{\dot{\bm{r}}^{\, 2}(u)}{4D} - V(\bm{r}(u),u) \right)\,.
\end{align} 
Note that Eq.~\eqref{eq:path-integral-solution} indicates that evolutionary paths passing through regions with high vaccine fitness tend to have larger weights in the path integral expression.  

Equation~\eqref{eq:path-integral-solution} presents a formal expression for the solution to the mean-field dynamics, applicable to any arbitrary initial distribution and fitness landscape. For the optimization of bnAbs, we substitute the germline distribution~\eqref{eq:n0-def} into Eq.~\eqref{eq:path-integral-solution}  and then apply the linear coordinate transformation    
$\bm{x} \to \bm{x} + \bm{x}^*(t)$, where
\begin{align}   \label{eq:xstar-def}
    \bm{x}^*(t) \equiv  
    \frac{\bm{\mu}_0}{1+\frac{\sigma_0^2}{2Dt_f}} \left(1-\frac{t}{t_f}\right) \,.  
\end{align}
After some algebra as described in Appendix~\ref{appendix:path-integral}, and upon comparing the resulting expression with the original one, we derive the following exact relationship
\begin{align}   \label{eq:nfx0-ratio-formula}
    \dfrac{n_f(0) \big\rvert_{\bm{\mu}_0 = \bm{\ell} ; V}}{n_f(0) \big\rvert_{ \bm{\mu}_0 = 0;\widetilde{V} }} 
    = 
    \exp{ - \frac{\bm{\ell}^2}{2(\sigma_0^2 + 2Dt_f)} } \, .
\end{align}
In this equation, the numerator on the left-hand side denotes the final population at $\bm{x}=0$ (i.e., the number of final bnAbs) given that the normally-distributed initial population~\eqref{eq:n0-def} is centered at $\bm{\mu}_0 =\bm{\ell}$ and evolves under the vaccine fitness $V(\bm{x},t)$.  
The denominator, on the other hand, represents the final bnAb count for a germline population that is centered at $\bm{\mu}_0 = 0$ and is evolved under the transformed fitness $\widetilde{V}$ defined as 
\begin{align}   \label{eq:V-tilde-def}
    \widetilde{V}(\bm{x},t) \equiv V \left( \bm{x}+\bm{x}^*(t),t \right) \, .
\end{align}
Therefore, the exact relationship~\eqref{eq:nfx0-ratio-formula} simply relates the number of bnAbs generated by fitness $V$, with the germline population initially centered away from the target in  shape space, to the number of bnAbs developed under the modified fitness $\widetilde{V}$, assuming the germline is already centered at the target (see top panel of Fig.~\ref{fig:PI-graphical}).

The crucial observation here is that the right-hand side of Eq.~\eqref{eq:nfx0-ratio-formula} is independent of the choice of the vaccine parameters. Thus, the maximum of the numerator on the left-hand side as $V$ is varied (by changing $\{ \bm{x}_V(t),\sigma_V(t) \}$), must coincide with the maximum of the denominator as $\widetilde{V}$ is varied. 
As a result, Eq.~\eqref{eq:nfx0-ratio-formula} effectively converts the maximization of 
$n_f(0)\big\rvert_{\bm{\mu}_0=\bm{\ell} ; V}$ into the maximization of $n_f(0) \big\rvert_{\bm{\mu}_0 = 0; \widetilde{V}}$. 
This mapping simplifies the optimization problem:
since for $\bm{\mu}_0 = 0$, the starting germline population is already  peaked at, and symmetric around, the target bin  $\bm{x}=0$, it makes sense that for $\widetilde{V}$ to be optimal, it should also be peaked at the target and symmetric with respect to it at all times (see Fig.~\ref{fig:PI-graphical}). 
Using this argument, and recalling that the maxima of the numerator and denominator in Eq.~\eqref{eq:nfx0-ratio-formula} coincide, we then conclude that
\begin{equation}   \label{eq:xv-optimal-xstar} 
    \bm{x}_V(t) \big\rvert_{\text{optimal}} = \bm{x}^*(t) 
    =\bm{\mu}_0~\frac{2D(t_f-t)}{{2Dt_f}+{\sigma_0^2}} 
    \, .
\end{equation}

It is worth noting that for a narrow germline population ($\sigma_0 \ll \sqrt{Dt_f}$), the optimal vaccine center given by Eq.~\eqref{eq:xstar-def} simply starts close to the naive population at $\bm{\mu}_0$ and reaches the target state ($\bm{x}=0$) at the final time. 
However, for a wider initial population ($\sigma_0 \gtrsim \sqrt{D t_f}$), the optimal center is shifted closer to the target, reflecting a change in the importance of paths contributing to $n_f(0)$ in Eq.~\eqref{eq:path-integral-solution}.  
Intuitively, one needs to position the vaccine fitness over time such that it gradually moves the germline population from $\bm{x}=\bm{\mu}_0$ towards the target state at $\bm{x}=0$, while also remaining sufficiently close to the peak of the existing B cell population to avoid  extinction by apoptosis. 
If $V(\bm{x},t)$ is positioned too close to the population $n(\bm{x},t)$, it may not optimally move the population to reach $\bm{x}=0$ within the finite time interval $t_f$; conversely, if $V$ is centered too far from the peak of $n(\bm{x},t)$, there is a risk of population collapse as the  vaccine-induced replication is directed to locations with fewer B cells to begin with. 
The optimal center (Eq. ~\eqref{eq:xstar-def}) aligns with this intuitive understanding and also demonstrates how the germline width $\sigma_0$ influences the optimal position of the vaccine fitness. 
In addition, Eq.~\eqref{eq:xv-optimal-xstar} reveals that the mean-field optimal center moves with a constant velocity from its starting position in the shape space towards the target bin. 
These features are further illustrated in the numerical and simulation results discussed in Section~\ref{sec:computational} and Appendix~\ref{appendix:trajectory}.

%% file: 3B_operator-approximation.tex
\subsection{Operator approximation and the optimal spread of the fitness function imposed by vaccine antigens}    \label{sec:bch}

As discussed above, Eq.~\eqref{eq:nfx0-ratio-formula} maps the problem of maximizing bnAbs to finding the optimal modified fitness function  
$\widetilde{V}$ that maximizes 
$n_f(0) \big\rvert_{\bm{\mu}_0 = 0 ; \widetilde{V}}$.  
Upon replacing $V$ by $\widetilde{V}$, Eq.~\eqref{eq:mean-field}    
describes the expansion of the transformed germline population centered at $\bm{x}=0$ under the modified fitness $\widetilde{V}$, coupled with diffusive spreading that arises from BCR mutations. 

In order to investigate the optimal vaccine spread, we employ operator approximations. 
Starting from Eq.~\eqref{eq:trotter}, we rearrange the operators in the time-ordered product formula  using a second-order truncation of the celebrated Baker-Campbell-Hausdorff  (BCH) formula~\cite{suzuki1977}, namely  
\begin{equation} \label{eq:BCH-def}
    e^{\epsilon A} e^{\epsilon B} = e^{\epsilon B} e^{\epsilon A} e^{\epsilon^2[A,B]} \left(1+O(\epsilon^3)\right),
\end{equation}
where $A$ and $B$ are generally non-commuting operators, and $C \equiv [A,B] = AB - BA$. 
In this formula, we assign 
$\epsilon \to \delta t$, 
$A \to D\nabla^2$, and 
$B \to \widetilde{V}_j (\bm{x})$,  
and then utilize this formula to reorder the alternating operator product in Eq.~\eqref{eq:trotter} such that all vaccine fitness operators $e^{\delta t \,\widetilde{V}_j}$ are grouped on the left in the product, all the diffusion operators $e^{\delta t D \nabla^2}$ are collected on the right, and the commutator terms that result from swapping these operators appear in between. The apoptosis operator $e^{-\lambda t_f}$ commutes with all the other operators.

To achieve such an ordering of the operators, we start by moving $e^{\delta t \widetilde{V}_0}$ through the $M$ diffusion operators $e^{\delta t D \nabla^2}$ on its left in Eq.~\eqref{eq:trotter}, resulting in a net commutator term 
$e^{(\delta t)^2 M C_0}$, where the commutator operator reads 
\begin{equation}    \label{eq:commutator-def}
    C_0 = [D\nabla^2 , \widetilde{V}_0(\bm{x})] = 2 D \widetilde{V}_0'(\bm{x}) \nabla + D\widetilde{V}_0''(\bm{x}) \, .
\end{equation}
Subsequently, the next fitness operator, $e^{\delta t \widetilde{V}_1}$, needs to pass through $(M-1)$ diffusion operators on its left, thereby creating 
a commutator contribution $e^{(\delta t)^2 (M-1) C_1}$. 
This process continues for all fitness operators, and in the limit of small increments $\delta t = \frac{t_f}{M} \to 0$, we obtain\footnote{Each truncation of the BCH formula introduces an error of order $(\delta t)^3$. It can be shown that the number of such error terms grows at least as $\sim M^3$, rendering this truncation an uncontrolled approximation. 
Despite this, the result obtained through this approach matches that of a perturbative solution in powers of the non-dimensionalized diffusion coefficient.   
The second-order truncation remains valid for small diffusion limits and as long as the total time $t_f$ is sufficiently short.}
\begin{align} \label{eq:BCH_nfinal}
    n_f  \big\rvert_{\bm{\mu}_0=0;\widetilde{V}}
    \approx & N_0 \quad e^{-\lambda t_f} e^{\int_0^{t_f} dt \, \widetilde{V}(\bm{x},t)} 
    \nonumber \\ 
    & \times e^{D\int_0^{t_f} dt \, (t_f - t) \, \left( 2\widetilde{V}'(\bm{x},t)\nabla + \widetilde{V}''(\bm{x},t)\right)}
    \nonumber \\ 
    & \times e^{t_f D \nabla^2} \, \mathcal{G}(\bm{x}_0,\sigma_0^2) \, .
\end{align}
Unlike the exact expression in Eq.~\eqref{eq:trotter}, where the fitness and mutation (diffusion) alternate in the operator product, Eq.~\eqref{eq:BCH_nfinal} suggests that $n_f$ can be approximated by first diffusing the starting profile $n_0$ under the (full) diffusion operator $e^{t_f D\nabla^2}$. Then, the commutator operators act on this diffused population profile, and, finally, the vaccine fitness operators (and the death operator) expand the profile to its final form. 
The action of $e^{t_f D\nabla^2}$ on the Gaussian initial population  widens its profile and yields 
$\mathcal{G}(\bm{x}_0,\sigma_0^2 + 2Dt_f)$, which is independent of the fitness profile. 
The symmetry of this updated profile with respect to $\bm{x}_0 = 0$ still implies a similar symmetry for the optimal
$\widetilde{V}(\bm{x},t)$, resulting in $\widetilde{V}'(\bm{x}=0,t) = 0$. Consequently, Eq.~\eqref{eq:BCH_nfinal} simplifies to  
\begin{align}   \label{eq:nfx-BCH-simplified}
    n_f(0) \big\rvert_{\bm{\mu}_0=0;\widetilde{V}} \approx \quad & 
    N_0 e^{-\lambda t_f} 
    e^{\int_0^{t_f}dt \widetilde{V}(0,t)}  \\
    \times 
    &e^{D \int_0^{t_f} dt (t_f-t) \widetilde{V}''(0,t) } \,
    \mathcal{G}(0,\sigma_0^2 + 2Dt_f) \, . \nonumber
\end{align}

For any choice of the vaccine-induced fitness profile, Eq.~\eqref{eq:nfx-BCH-simplified} offers a starting point to investigate the optimal vaccine spread by providing an explicit expression whose optima can be derived by straightforward differentiation.    
Narrowing our focus, we consider the  Gaussian vaccine-induced fitness
\begin{align}   \label{eq:Vg-def}
    V_\mathcal{G}(\bm{x},t) \equiv A_V \, \mathcal{G} \left(\bm{x}-\bm{x}_V(t),\sigma_V(t)^2 \right) \, .
\end{align} 
Recall that this choice is consistent with our initial assumption that, in the shape-space representation, the fitness can be parameterized by the location of its center and its spread. Furthermore, Eq.~\eqref{eq:Vg-def} also satisfies the total-dose constraint given by Eq.~\eqref{eq:constraint}. 
However, it is important to note that Eq.~\eqref{eq:Vg-def} does not represent the most general Gaussian fitness profile allowed by the constraint, as it only considers a fixed \textit{vaccine strength} $A_V$.  
The more general case would involve a vaccine dose that varies over time, potentially leading to larger and more diverse antibody responses~\cite{tam-2016sustained,cirelli-2019slow,leerang-bioarxiv}.
We defer the investigation of such scenarios to future studies. 

By selecting the optimal vaccine center according to Eq.~\eqref{eq:xv-optimal-xstar}, the Gaussian fitness of Eq.~\eqref{eq:Vg-def} transforms according to Eq.~\eqref{eq:V-tilde-def}, resulting in the modified fitness
$\widetilde{V}_{\mathcal{G}}(\bm{x},t) = A_V \mathcal{G}(\bm{x},\sigma_V^2(t))$, which is still a Gaussian profile but now centered at the origin. 
Substituting $\widetilde{V}_{\mathcal{G}}$ into Eq.~\eqref{eq:nfx-BCH-simplified}, we obtain
\begin{align}
     &n_f(0) \big\rvert_{\bm{\mu}_0=0;\widetilde{V}_\mathcal{G}}\approx 
     \frac{N_0 e^{-\lambda t_f}}{\sqrt{2\pi(\sigma_0^2+2Dt_f)}}
     \\
     &\qquad \times \exp{\int_0^{t_f} dt \, \left( \frac{A_V}{\sqrt{2\pi} \sigma_V(t)} \left(1-\frac{D(t_f-t)}{\sigma_V^2(t)}\right) \right)} \, .
     \nonumber
\end{align}
To maximize this expression, we set its functional derivative with respect to $\sigma_V$ to zero. Assuming $\sigma_V(t) \geq \sigma_{\min}$ at all times (see the discussion below), we finally obtain
\begin{align}   \label{eq:sigma-bch}
    \sigma_{\mathrm{BCH}}(t) = \mathrm{max} \left\{ \sqrt{ 3 D (t_f - t)} \, , \sigma_{\min}   \right\}
\end{align}
In deriving Eq.~\eqref{eq:sigma-bch}, we have assumed that $\sigma_V(t) \geq \sigma_{\min}$, which prevents $V_{\mathcal{G}}$ from reaching arbitrarily large values by imposing a maximum fitness value 
$A_V/(\sqrt{2\pi} \sigma_{\min})$.  
As discussed in Section~\ref{sec:model}, the fitness function exhibits a minimal yet finite spread for vaccine shots containing a single antigen type, whereas cocktail shots typically exhibit a wider spread. 
Moreover, in Appendix~\ref{appendix:bch-discrete}, we demonstrate that the optimal spread retains a finite minimum value when employing the BCH approximation for B cell dynamics on discrete similarity bins (before taking the continuum limit in shape space). Additionally, the minimum spread can also reflect practical constraints on how precisely  the vaccine-induced fitness can be concentrated in different regions of  shape space. We posit that $\sigma_{\min}$ encompasses all of these factors. 

Equation~\eqref{eq:sigma-bch} can be understood by noting that within a temporal window of duration $\Delta t$, diffusion affects a range of order $\sqrt{D \Delta t}$ of bins on the $x$ axis. To maximize $n_f(0)$ while adhering to a constrained total dose, it would thus be sensible to concentrate the fitness on BCRs in a region around the target bin that has the potential to reach the target location by mutation within the available time. 
Replication events at larger mutational distances are unlikely to reach the target within the finite allotted time.
This  suggests that at any time $0<t<t_f$, the optimal $\widetilde{V}$ should at most cover BCRs within a range of order $\sqrt{D(t_f - t)}$  while remaining peaked at $\bm{x}=0$. 
Once the vaccine spread reaches the smallest possible value $\sigma_{\min}$, it should remain at that value until the final time. 
In a qualitative sense, Eq.~\eqref{eq:sigma-bch} suggests that an optimal vaccination protocol should initially have a broader spread and contain shots with more diverse antigens to cover a larger range of BCRs, which would provide the germline population a chance to grow and diversify.  
As time progresses, vaccine epitopes gradually narrow down according to Eq.~\eqref{eq:sigma-bch}, concentrating the fitness on expanding those BCRs within an accessible mutational distance from the target bin.

It is also worth noting that the right-hand side of the BCH approximate result~\eqref{eq:nfx-BCH-simplified} depends only on time integrals of $\widetilde{V}(0,t)$ and $\widetilde{V}''(0,t)$, making it a strictly local function of the transformed vaccine fitness $\widetilde{V}$ at $\bm{x}=0$. 
This locality aligns with the second-order truncation of the BCH formula, as higher-order terms in BCH's nested commutator expansion would introduce contributions involving more derivatives of the fitness, leading to non-local expressions in $\bm{x}$.   
Since the mean-field Eq.~\eqref{eq:mean-field} aims to describe the coarse-grained temporal evolution of the B cells toward the bnAb state, and considering that bnAbs are typically produced over long timescales in natural circumstances~\cite{bnab-prevalence-2014,kreer-bnab_prob-2023}, it may be reasonable to assume the slow diffusion regime, where local approximations such as the one above remain reliable at the mean-field level.  
However, it is important to acknowledge that even in the slow diffusion limit, this approximation might break down if either the vaccine fitness or the B cell population profile is sharply peaked, as higher-order derivatives may become significant in the expansion.

Furthermore, as noted in Section~\ref{sec:mean-field-model}, our approach differs from Refs.~\cite{ganti-bnab-2021,kayla-bnab-2020} in our assumption that antigens administered at each time point share sufficient similarity,  that bnAb sequences are not inherently the fittest. 
Given this distinction and the discussion of the previous paragraph, while Eq.~\eqref{eq:sigma-bch} bears resemblance to the findings of Ref.~\cite{ganti-bnab-2021}-- where the optimal fitness function appeared narrower in subsequent vaccinations--we approach the interpretation of Eq.~\eqref{eq:sigma-bch} with caution.  
In the next section, we will compare the performance of protocols employing a width $\sigma_{\mathrm{BCH}}$ with those utilizing a narrow fitness width $\sigma_{\min}$ through numerical results and simulations. The comparative analysis aims to shed light on the effectiveness of different vaccine spread strategies and will elucidate a distinct interpretation of the optimal spread in our study.

%% file: 4_computational.tex
\begin{figure*}[t]
 \begin{minipage}[c]{.45\linewidth}
		\centering
		\includegraphics[width=\linewidth]{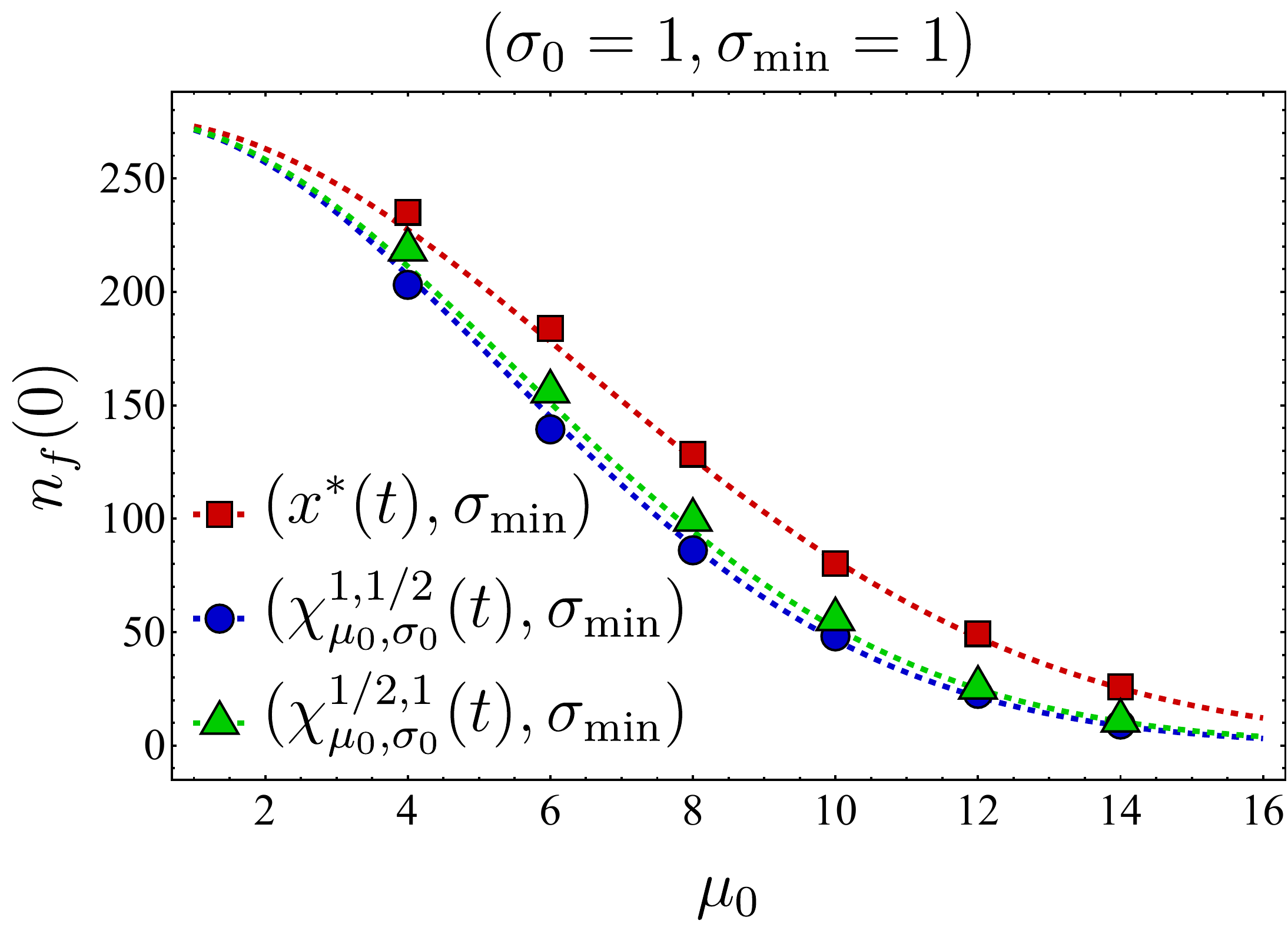}
	\end{minipage} 
	\hskip.5cm
 \begin{minipage}[c]{.45\linewidth}
		\centering
		\includegraphics[width=\linewidth]{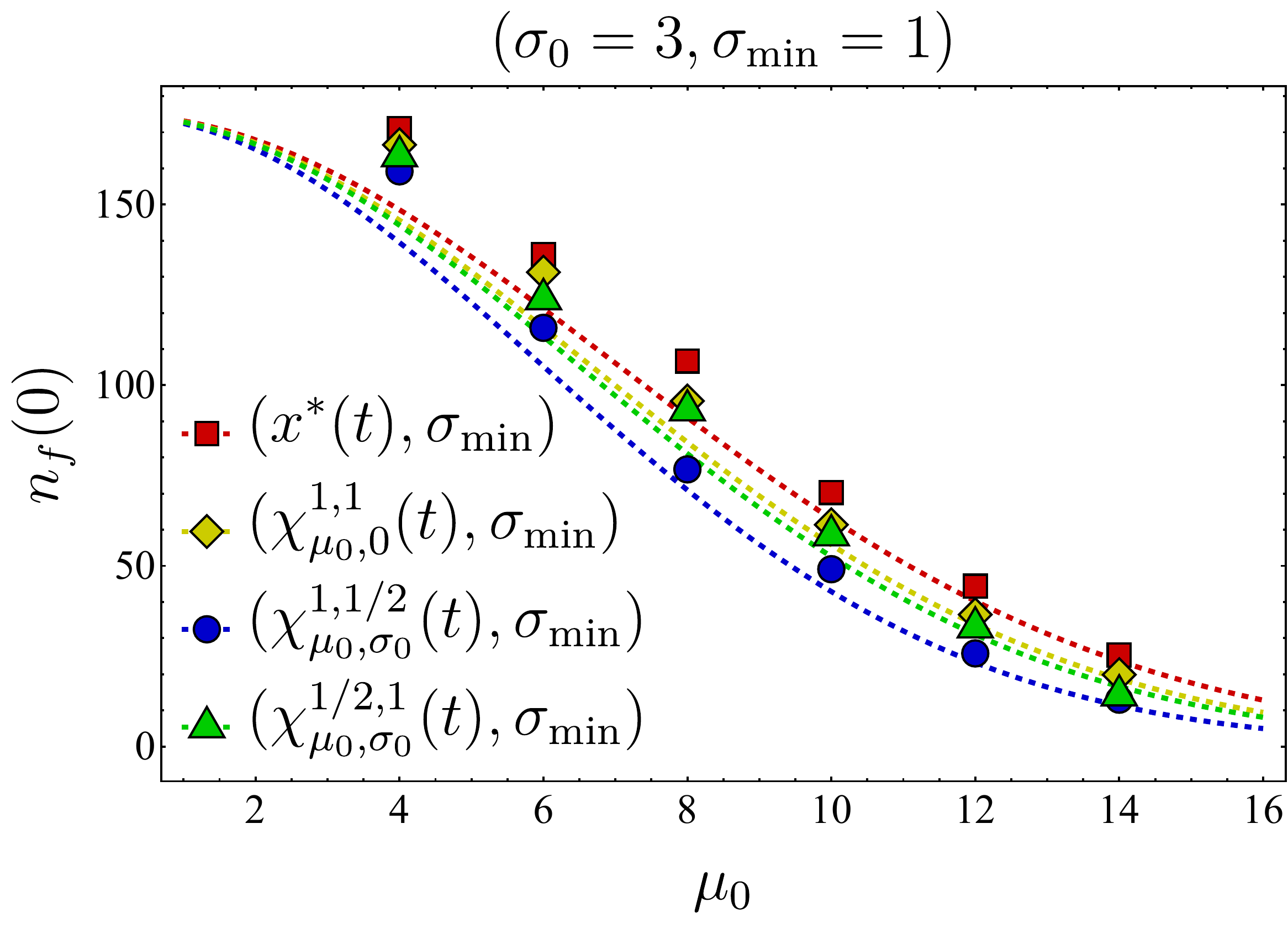}
	\end{minipage} 
\caption{
Final bnAb counts for different vaccination protocols in a one-dimensional shape space as obtained from the continuum mean-field numerics (dashed lines) and SSA averages (solid points), with the choice of parameters $A_V = 0.08$, $D=0.05$, $\lambda=0.001$, and $t_f = 400$. 
Left: final bnAb count as a function of the germline center $\mu_0$ assuming a fixed germline width $\sigma_0=1$. 
The vaccine-induced fitness is assumed to follow the Gaussian form of Eq.~\eqref{eq:Vg-def} with the vaccine center located at $\chi^{q,p}_{\mu_0,\sigma_0}(t)$ (see Eq.~\eqref{eq:chi-def}) and the vaccine spread set to the minimum allowed
$\sigma_{\mathrm{min}}=1$. Different colors correspond to different choices for $q$ and $p$, with $q=p=1$ corresponding to the optimal center $x^*(t)$. The $x^*(t)$ protocol indeed outperforms the other choices for the vaccine center. 
Right: similar to the left panel but for a wider germline width ($\sigma_0=3$). We have also included the results obtained from setting the vaccine center according to $\chi^{1,1}_{\mu_0,0}(t)$. Here, SSA averages deviate from the continuum mean-field results for $\mu_0 \sim \sigma_0$, indicating the breakdown of the continuum approximation. 
}
   \label{fig:nf0-vs-mu0}
\end{figure*}

In this section, we delve into the dynamics of the B cell population within a one-dimensional shape space ($d=1$) and investigate both the numerical solution of the continuum mean-field model (Eq.~\eqref{eq:mean-field}) and simulations of the discrete stochastic dynamics (Eq.~\eqref{eq:master-fock}). 
It is important to recall that while the master equation deals with discrete shape-space bins, representing BCR occupancy with integer values, the mean-field model offers a continuum description, neglecting population fluctuations. Consequently, it is not immediately clear whether discrete sample averages would align with the mean-field results. 
To establish a comparison between the two approaches, we employ the  stochastic simulation algorithm (SSA) \textit{a la} Gillespie~\cite{gillespie-review-2007,gillespie-1977} to generate realizations of stochastic dynamics and then compare the sample averages with those obtained from the mean-field treatment. 
This numerical examination sheds light on how effectively the optimal vaccine fitness obtained from the mean-field model translates to the realm of discrete stochastic dynamics

To set up our numerical experiments and simulations, we must define parameters for:  
(1) the germline population, specifically $N_0$, $\mu_0$, and $\sigma_0$; 
(2) the dynamics, including $D$, $t_f$, $\lambda$; and
(3) the vaccine's strength, denoted by $A_V$. 
To make these parameter choices more grounded, we draw from empirical observations and rough estimates. 
First, it is cruical to consider the slow and delayed production of bnAbs in HIV patients~\cite{bnab-prevalence-2014}. This implies that the germline population should be initially distant from the target bin ($x=0$). For instance, it is known that numerous mutations are typically required for germline B cells to evolve into bnAbs that can target CD4 binding sites~\cite{burton-review-2016}. This informs a constraint  among our parameters: 
\begin{align}   \label{eq:mu0-condition}
    \mu_0 \gtrsim \sqrt{Dt_f}.
\end{align}

Next, in the absence of vaccine-induced replication, the apoptotic B cell population is expected to go extinct before reaching the bnAb state through mutations. In a mean-field approximation, we crudely define extinction as a total population size falling below $1$. This extinction condition can be expressed as   
\begin{equation}    \label{eq:N0-condition}
    \ln N_0 \lesssim \frac{{\lambda} \mu_0^2}{D}. 
\end{equation} 

Drawing from the findings of Ref.~\cite{ganti-bnab-2021}, we posit that an effective vaccine would provide sufficient fitness to prevent the extinction.  
We thus choose the vaccine strength $A_V$ so that the total vaccine-induced fitness given at each time exceeds the total death rate across all shape-space bins, i.e., 
\begin{equation}    \label{eq:Av-condition}
    A_V \gtrsim \lambda N_{\text{bin}} \Delta x_{\text{bin}},
\end{equation}
where $N_{\text{bin}}$ represents the number of bins and $\Delta x_{\text{bin}}$ is the length of a bin, assumed to be one. (Note that as per Eq.~\eqref{eq:Vg-def}, $A_V$ has the dimension of (shape-space) length per time.)

Lastly, we note that a further condition to avoid extinction under the optimal vaccine center is for $x^*$ to remain close to the population's center; otherwise, most of the provided fitness is expended on bins with few to no B cells.  
Given the difficulty of analytically solving for the population center as a function of time, we only impose this condition on the germline at $t=0$, which corresponds to    
$|\mu_0 - x^*(t=0)| \lesssim \sigma_0$. 
Substituting for $x^*(t=0)$ from Eq.~\eqref{eq:xstar-def} then yields the fourth condition: 
\begin{equation}
    t_f \gtrsim \frac{\sigma_0 (\mu_0 - \sigma_0)}{D},
\end{equation}
which, for a given diffusion constant $D$, sets a lower bound for the total vaccination time $t_f$. Typically $\sigma_0 \ll \mu_0$, ensuring that this bound is significantly smaller than the diffusive timescale ${\mu_0^2}/{D}$ required for the germline to reach the target by mutations alone (c.f. Eq.~\ref{eq:mu0-condition}).  

Guided by these conditions, we choose  $N_{\text{bin}} = 43$ corresponding to the shape-space bins $ x \in  \{ -21,-20,\ldots,0,\ldots,20,21\}$ for the stochastic simulations. This range is associated with the interval $(-21,21)$ for the continuum mean-field numerics.  We set the death rate to  $\lambda = 0.001$ and the mutation rate to $\mu=0.1$ in SSA, corresponding to a mean-field diffusion coefficient of $D=0.05$ (see Appendix~\ref{appendix:master-eq}). 
We assume the germline population is normally distributed according to Eq.~\eqref{eq:n0-def}, with an initial population size of $N_0=10$. In addition, the imposed vaccine fitness follows the Gaussian shape of Eq.~\eqref{eq:Vg-def}, where we fix the vaccine strength at $A_V=0.08$.  
For the stochastic dynamics, we utilize SSA to take averages over $1000$ realizations of  Eq.~\eqref{eq:master-fock} for a total time of $t_f = 400$. To facilitate the simulation, we implement SSA along with an operator splitting scheme~\cite{operator-splitting} where we alternate between periods of stochastic replication events and stochastic mutation or death events (cf. Eq.~\eqref{eq:BCH-def} for the corresponding mean-field expression). In the following, we present the SSA results for $M=200$ steps, although consistent results are obtained  for $M=400$. Note that the mean-field numerics presented below are obtained by direct numerical solution of the continuous-time dynamics described given by Eq.~\eqref{eq:mean-field}. 

Fixing the vaccine spread at $\sigma_{\min}=1$ and the germline width at $\sigma_0=1$, we first examine the effect of different vaccine centers on the final bnAb count by selecting the vaccine center from the family 
\begin{align}   \label{eq:chi-def}
    \chi^{q,p}_{\ell,\kappa}(t) \equiv \frac{\ell}{1+\frac{\kappa^2}{2Dt_f}}\left(1-\left(\frac{t}{t_f}\right)^q \right)^p \, . 
\end{align}
In particular, we compare the final bnAb count for $\chi^{1,1}_{\mu_0,\sigma_0}$-- corresponding to $x^*$ as defined in Eq.~\eqref{eq:xstar-def}--with that of $\chi^{1,1/2}_{\mu_0,\sigma_0}$ (vaccine center accelerating towards the target), $\chi^{1/2,1}_{\mu_0,\sigma_0}$ (vaccine center decelerating towards the target), and $\chi^{1,1}_{\mu_0,0}$ (uniform motion similar to $x^*$ but not accounting for germline's width). 
The left panel in Fig.~\ref{fig:nf0-vs-mu0} shows the final bnAb counts as a function of the germline distance from the target for three different vaccine centers $x^*(t)$ (red), $\chi^{1,1/2}_{\mu_0,\sigma_0}(t)$ (blue), and $\chi^{1/2,1}_{\mu_0,\sigma_0}(t)$ (green). 
We observe that in each case, the SSA sample averages (solid points) closely match the continuum mean-field numerics (dashed lines). 
To understand this, note that with $\sigma_0 = 1$, the germline population is narrowly distributed, so the entire population would roughly move together in  shape space. When this population is relatively far from $x=0$, moving between neighboring (discrete) bins would only change the distance to the target bin incrementally, thus making the continuum limit a reliable approximation of the discrete dynamics. Additionally, with the chosen vaccine strength (Eq.~\eqref{eq:Av-condition}), the population size increases and makes the population number fluctuations less relevant, thus rendering the mean-field description even more accurate. 
This is also reflected in the fact that the optimal center $x^*(t)$--obtained from the continuum mean-field approximation--outperforms the other protocols even in the discrete stochastic dynamics. 

The right panel in Fig.~\ref{fig:nf0-vs-mu0} presents a similar comparison of the final bnAb counts, but for a broader germline with $\sigma_0 = 3$. Here, we include a fourth choice for the vaccine center, namely $\chi^{1,1}_{\mu_0,0}(t)$ (in yellow), which differs from $x^*(t)$ in that its starting position and velocity are not adjusted by $\sigma_0$. We still observe that $x^*(t)$ outperforms the other protocols, although the final bnAb counts (vertical axis) are generally smaller compared to the $\sigma_0 = 1$ case. This is because the population does not move toward the target as synchronously when the initial population is spread more widely. 
It is also seen that when $\mu_0$ becomes comparable with $\sigma_0$, SSA averages deviate from the continuum mean-field results, which might indicate the breakdown of the continuum approximation for that range of parameters. Nevertheless, the qualitative trend of the SSA sample averages is still captured by the mean-field solution. 
In Appendix~\ref{appendix:trajectory}, we present temporal snapshots of the BCR population evolved under various vaccine protocols and explore additional aspects of the population dynamics, such as population size and the location of the population peak over time. Notably, these observations indicate that, as discussed below Eq.~\eqref{eq:xv-optimal-xstar}, the optimal center gradually shifts the germline population from its initial position in  shape space toward the target while maintaining an optimal distance from it. 

\begin{figure}[t]
   \begin{minipage}[c]{.8\linewidth}
		\centering
		\includegraphics[width=\linewidth]{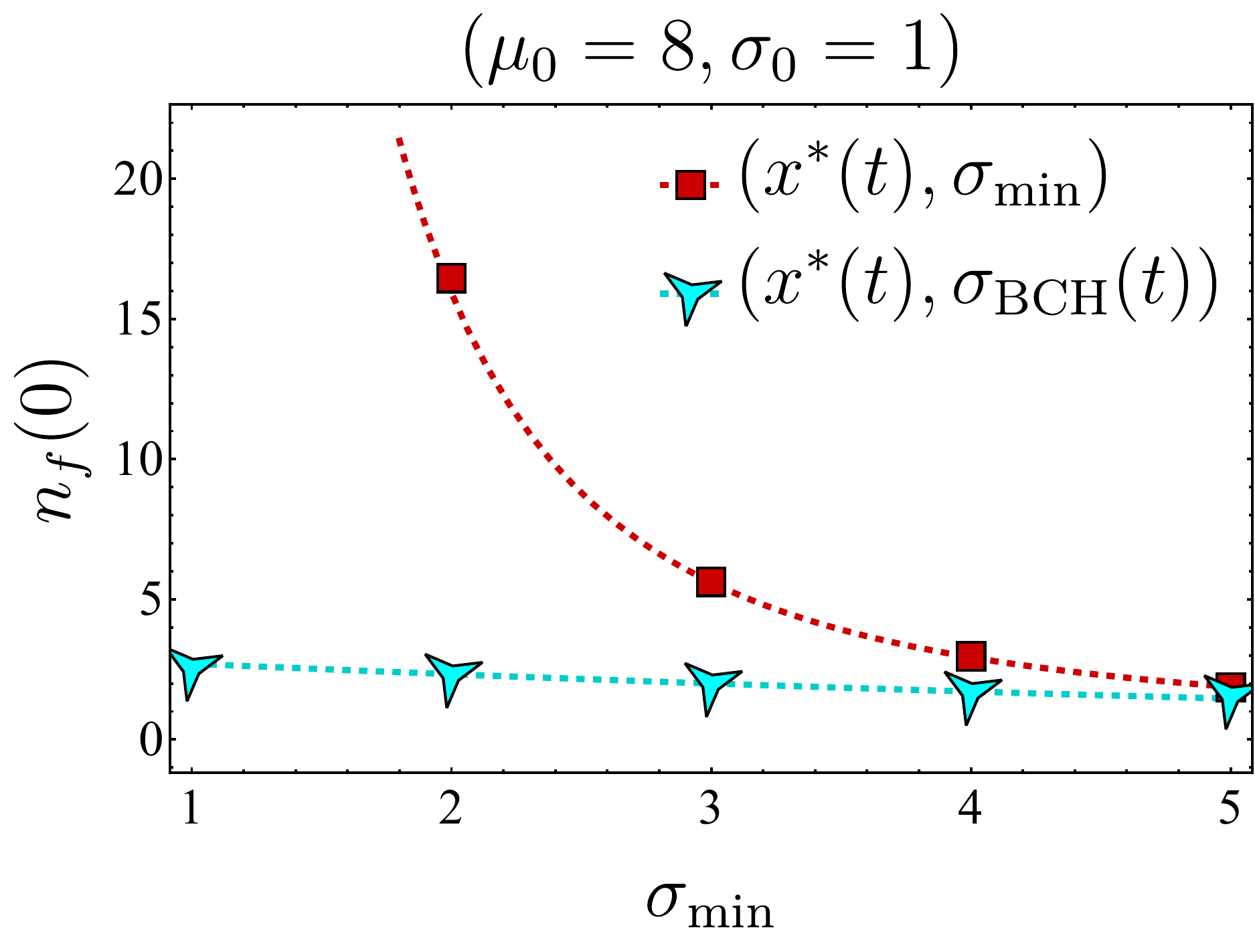}
	\end{minipage} 
\caption{ 
Final bnAb numbers for a germline centered at $\mu_0=8$ and with width $\sigma_0 = 1$ in the one-dimensional shape space. The population is evolved under two different vaccine fitnesses,  both optimally centered at $x^*(t)$ but with different spreads. 
The dashed lines represent results from continuum mean-field numerics, while the solid points depict SSA averages [with the choice of parameters $A_V = 0.08$, $D=0.05$, $\lambda=0.001$, and $t_f = 400$].
This plot illustrates that maintaining the vaccine spread at the minimum allowed ($\sigma_{\min}$) yields a larger bnAb count compared to using the time-dependent spread prescribed by the BCH result,~\eqref{eq:sigma-bch} and may indicate that a sequential vaccination with a single antigen type administered at a time would outperform a cocktail regimen. }
   \label{fig:nf0-vs-sigmamin}
\end{figure}

Assuming the optimal choice for the vaccine center, we proceed to compare the effectiveness of two different vaccine spread strategies: maintaining the fitness as narrow as possible or adjusting the fitness spread based on the time-dependent BCH result. 
Figure~\ref{fig:nf0-vs-sigmamin} depicts the final bnAb counts as a function of $\sigma_{\min}$. It shows that maintaining the spread at $\sigma_{\min}$ (in red) results in higher final bnAb numbers compared to setting the vaccine spread according to the time-dependent BCH result given in Eq.~\eqref{eq:sigma-bch} (in cyan). 
This observation suggests that maximizing the focus of fitness while aligning its center with the optimal position (as defined in Eq.~\eqref{eq:xstar-def}) results in the largest bnAb numbers ultimately. 
Drawing from the discussion on the minimum fitness spread in Section~\ref{sec:bch} (also see Section~\ref{sec:minimal-model}), a plausible interpretation of this finding suggests that sequential immunization with a single antigen type, administered at any time, outperforms immunization with cocktails of antigens for which the fitness spread is larger, in agreement with the findings of Ref.~\cite{wang-bnab-2015}. This observation implies that even a cocktail that optimally `frustrates' the B cell population, as found in Ref.~\cite{shaffer2016optimal}, would be outperformed by a sequential immunization regimen containing appropriately selected single antigen types at each time point. 

It is also noteworthy to contrast Fig.~\ref{fig:nf0-vs-sigmamin}  with the conclusions drawn in Ref.~\cite{ganti-bnab-2021,kayla-bnab-2020}, where the administered vaccine antigens were considered sufficiently distinct in their variable residues such that the fittest BCR sequence corresponded to bnAbs, while the germline would start far from the bnAb sequence. In Ref.~\cite{ganti-bnab-2021}, the optimal spread of the vaccine-induced fitness function was found to be one that is sufficiently wide to reach the least cross-reactive existing B cells; a narrower spread would have led to B cell extinction. Consequently, the optimal fitness landscape narrowed in subsequent immunizations, a phenomenon that coincides with the narrowing nature of the BCH result (Eq.~\ref{eq:sigma-bch}). As discussed in the preceding paragraph, the results presented in this work relax the assumption of distinct antigens and instead concern vaccine shots containing similar antigen types.  


%% file: 5_discussion.tex
We have explored the optimal vaccination protocol for producing bnAbs through a minimal birth-death-mutation model described by the master equation (Eq.~\eqref{eq:master-fock}) and its continuum mean-field approximation (Eq.~\eqref{eq:mean-field}).  
Assuming a B cell germline that is initially distant from the desired bnAb state in the shape space, this optimization problem comprises two aspects:  first, determining the optimal time-dependent center of the vaccine-induced fitness function, and, second, its optimal time-dependent spread. 
The center of the vaccine fitness function--i.e., where the fitness peaks--is determined by the epitopes shared among the antigen(s) presented at each time point; the spread of the fitness function, on the other hand,  is controlled by diversity of the presented epitopes in each shot.   
We assumed the administered antigens are sufficiently similar to allow for approximating the vaccine-induced fitness landscape with a coarse-grained smooth function whose peak can be distal from the BCR sequence corresponding to bnAbs (Fig.~\ref{fig:schematic}). 


We addressed the first part by employing the path integral formulation of the mean-field dynamics (Eq.~\eqref{eq:path-integral-solution}) and used a linear coordinate transformation to derive a nontrivial exact relationship given in Eq.~\eqref{eq:nfx0-ratio-formula}. This relationship is valid for a general vaccine-induced fitness if the germline population is normally distributed (cf. Eq.~\eqref{eq:n0-def}). 
Equation~\eqref{eq:nfx0-ratio-formula} maps the maximization of the final bnAb count $n_f(0)$ for a germline centered at $\bm{\mu}_0 \neq 0$ away from the target, into the simpler problem of finding the optimal transformed fitness, defined in Eq.~\eqref{eq:V-tilde-def}, that maximizes the bnAb number for a germline that is instead initially centered at the target ($\bm{\mu}_0=0$).  
Based on this mapping, we then argued (see Fig.~\ref{fig:PI-graphical}) that the optimal vaccine center follows Eq.~\eqref{eq:xv-optimal-xstar}, with $\bm{x}^*(t)$ defined in Eq.~\eqref{eq:xstar-def}. 

The same mapping also allows us to investigate the second part of the bnAb optimization, i.e., the optimal spread of the vaccine-induced fitness landscape. Using a second-order truncation of the celebrated BCH formula, we obtained an approximate form for the mean-field solution (Eq.~\eqref{eq:nfx-BCH-simplified}) that, given a specific form of fitness, can be used to examine how the vaccine spread affects the final bnAb count. In the special case of a Gaussian vaccine-induced fitness (Eq.~\eqref{eq:Vg-def}), maximizing this expression with respect to the vaccine spread led us to Eq.~\eqref{eq:sigma-bch}. In Eq.~\eqref{eq:sigma-bch}, we incorporate $\sigma_{\min}$ as a proxy to encompass all factors constraining the narrowness of the fitness function. These factors may include the (weaker) binding of neighboring BCRs to the presented antigen, the effect of discrete similarity bins, and  experimental limitations. 

Finally, to test the performance of the optimal fitness as obtained from the mean-field model in the discrete stochastic dynamics, we simulated Eq.~\eqref{eq:master-fock} using the Gillespie SSA and compared the resulting sample averages with the numerical solution of the associated mean-field dynamics. We observed that the SSA averages and mean-field numerics match closely, and the optimal vaccine center, Eq.~\eqref{eq:xv-optimal-xstar}, outperforms other protocols even at the level of the discrete stochastic dynamics (Fig.~\ref{fig:nf0-vs-mu0}). 
In addition, we found that keeping the vaccine spread at the minimum allowed seems to be optimal (Fig.~\ref{fig:nf0-vs-sigmamin}), suggesting that antigen cocktails reduce the effectiveness of the immunization procedure.   

Our findings suggest that optimal vaccination protocols could potentially be designed to maximize the evolution of rare bnAbs without risking B cell extinction by ensuring that sequentially administered vaccine antigens are not too different at each time point. This important conclusion of our work warrants further investigation by addressing several inherent limitations of the model discussed in Section~\ref{sec:model}, as well as accounting for continuous entry of naive B cells into germinal centers~\cite{hagglof2023continuous,victora-review-2022}.  As briefly discussed in Section~\ref{sec:model}, during affinity maturation, B cells interact with each other in various ways, such as competing for limited survival signals from T cells and memory effects including the epitope masking~\cite{akc-perspective-2017,leerang-omicron-2023}. Incorporating these effects in a birth-death-mutation model  introduces nonlinear terms, and may well lead to modified perspectives on bnAb evolution, compared to the linear model employed in this study. In  future, we intend to explore optimal vaccination strategies using a more realistic version of the stochastic dynamics that accounts for these crucial nonlinearities and feedback loops. Additionally, we plan to extend our analyses to encompass time-dependent vaccine dosages. Ultimately, we hope that our study will motivate further theoretical and experimental investigations by other researchers.

%% file: Appendix_master.tex
In this appendix, we elaborate on the derivation of the mean-field equation~\eqref{eq:mean-field} from the master equation~\eqref{eq:master-fock}. 
The average BCR occupation number at bin $\bm{x}$ and time $t$ is given by 
$\expval{n_{\bm{x}}(t)} = \sum_{\lbrace n \rbrace} n_{\bm{x}}  \, P(\lbrace n \rbrace , t)$, where the summation is performed over all possible population configurations $\lbrace n \rbrace$.  Note that due to the averaging process,  $\expval{n_{\bm{x}}(t)}$ is not necessarily integer-valued anymore, while $\bm{x}$ remains discrete at this stage.   
The dynamics of the mean population is obtained by taking a time derivative of $\expval{n_{\bm{x}}(t)}$. Using Eq.~\eqref{eq:master-fock}, it reads
\begin{align}
    \partial_t \expval{n_{\bm{x}}(t)} = 
    \quad &\sum_{\{n\}} n_{\bm{x}}\sum_{\bm{x}'} V_{\bm{x}'}(t) \, \big[ \mathbb{E}_{\bm{x}'}^{-}-1 \big] \big(n_{\bm{x}'} P(\lbrace n \rbrace ,t) \big)
    \nonumber\\
    \quad+&
    \sum_{\{n\}} n_{\bm{x}}\sum_{\bm{x}',\bm{y}'} \gamma_{\bm{x}' \bm{y}'} 
    \big[ \mathbb{E}^+_{\bm{x}'} \mathbb{E}_{\bm{y}'}^{-} -1 \big] 
    \big( n_{\bm{x}'} \, P(\lbrace n \rbrace , t) \big) \, 
    \nonumber\\
    \quad+&
    \sum_{\{n\}} n_{\bm{x}'}\sum_{\bm{x}'} \lambda_{\bm{x}'} \,
    \big[\mathbb{E}^+_{\bm{x}'} - 1\big] \big(n_{\bm{x}'}  P(\lbrace n \rbrace ,t) \big) \, , 
\end{align}
which can be simplified in each line by an appropriate shift of the variables in the sums over $\{n\}$. 
For instance, it is straightforward to show that  
\begin{align}   \label{eq:dt-expvaln}
&\sum_{\{n\}} n_{\bm{x}}\sum_{\bm{x}'} V_{\bm{x}'}(t) \, \mathbb{E}_{\bm{x}'}^- \left( n_{\bm{x}'} P(\{n\},t) \right) = \\
&\qquad \sum_{\{n\}}  \left(n_{\bm{x}} + \delta_{\bm{x},\bm{x}'} \right)\sum_{\bm{x}'} V_{\bm{x}'}(t) \, n_{\bm{x}'} P(\{n\},t) 
\end{align}
where $\delta_{\bm{x},\bm{x}'}$ denotes the Kronecker delta. On implementing these simplifications, Eq.~\eqref{eq:dt-expvaln} turns into the following mean-field dynamics 
\begin{align}   \label{eq:mean-field-discrete}
    \partial_t \expval{n_{\bm{x}}(t)} = &[V_{\bm{x}}(t) -\lambda_{\bm{x}}] \, \expval{n_{\bm{x}}(t)} \nonumber \\
    &+\sum_y [\gamma_{\bm{y}\bm{x}} \, \expval{n_{\bm{y}}(t)} - \gamma_{\bm{x}\bm{y}} \, \expval{n_{\bm{x}}(t)}] \, ,  
\end{align}
where $\bm{x}$ and $\bm{y}$ denote different affinity bins. 

To take the continuum limit of $\bm{x}$ in Eq.~\eqref{eq:mean-field-discrete}, we make the change 
\begin{align}
    &\sum_y [\gamma_{\bm{y}\bm{x}} \expval{n_{\bm{y}}} - \gamma_{\bm{x}\bm{y}}\expval{n_{\bm{x}}}] \to 
    \\
    &\int d^d\Delta \bm{x} \, \left[ \gamma(\bm{x}-\Delta \bm{x},\Delta \bm{x}) n(\bm{x}-\Delta \bm{x},t) - \gamma(\bm{x},\Delta \bm{x}) n(\bm{x},t) \right] \nonumber 
\end{align}
where $\gamma(\bm{x},\Delta \bm{x})$ denotes the mutation rate for a jump of length $\Delta \bm{x}$ that starts from $\bm{x}$.  
The assumption of localized jumps corresponds to negligible jump rates for large $\Delta \bm{x}$ (see the discussion is Section~\ref{sec:model}), allowing for a Taylor expansion of the integral term in powers of $\Delta \bm{x}$. Truncating the Taylor expansion at second order then gives  
\begin{align}   \label{eq:app-meanfield-full}
    \partial_t n(\bm{x},t) &\approx 
    \big[ V(\bm{x},t) - \lambda(x) \big] \, n(\bm{x},t) 
    \\ \nonumber
    &\quad + \nabla^2 \big[ D(\bm{x}) \, n(\bm{x},t) \big] - \nabla\cdot \big[\bm{w}(\bm{x}) \, n(\bm{x},t)], 
\end{align}
where we have defined the diffusion coefficient as
\begin{align}   \label{eq:mu-to-D}
    D(\bm{x}) = \frac{1}{2} \int d^d\Delta \bm{x} \, (\Delta \bm{x})^2 \, \gamma(\bm{x},\Delta \bm{x})
\end{align} 
and the drift velocity as $\bm{w}(\bm{x}) = \int d^d\Delta \bm{x} \, \Delta \bm{x} \, \gamma(\bm{x},\Delta \bm{x})$. 
For symmetric mutation rates, we have $\bm{w}(\bm{x},t)=0$. Assuming jump rates independent of $\bm{x}$ then simplifies Eq.~\eqref{eq:app-meanfield-full} into Eq.~\eqref{eq:mean-field} of the main text.

%% file: Appendix_path-integral.tex
 In this appendix, we provide details on the derivation of Eq.~\eqref{eq:nfx0-ratio-formula} from the path integral formula~\eqref{eq:path-integral-solution}. 
We first rewrite Eq.~\eqref{eq:path-integral-solution} in a new frame, where the origin moves according to
$\bm{z}(t) = \bm{z}_0 - \bm{v} t$.   
This transformation corresponds adjusting shape-space vectors as
$\bm{x} (t) \to \tilde{\bm{x}} = \bm{x}-\bm{z}(t)$ and paths as 
$\bm{r}(t) \to \tilde{\bm{r}}(t) = \bm{r}(t) - \bm{z}(t)$ while keeping the path integral measure unchanged 
($\mathcal{D}\tilde{\bm{r}}(t) = \mathcal{D}\bm{r}(t)$). 
In addition, the initial population transforms according to 
\begin{align*}
    &n_0(\bm{x}_0) = N_0 \mathcal{G}(\bm{x}_0 - \bm{\mu}_0, \sigma_0^2) 
    \\
    &\qquad\qquad\qquad \to \tilde{n}_0(\tilde{\bm{x}}_0) = N_0\mathcal{G}(\tilde{\bm{x}}_0 - \tilde{\bm{\mu}}_0,\sigma_0^2)
\end{align*}
where we have defined
$\tilde{\bm{\mu}}_0 \equiv \bm{\mu}_0 - \bm{z}(t=0) = \bm{\mu}_0 - \bm{z}_0$. 

\begin{figure}[t]
    \centering
    \includegraphics[width=.95\linewidth]{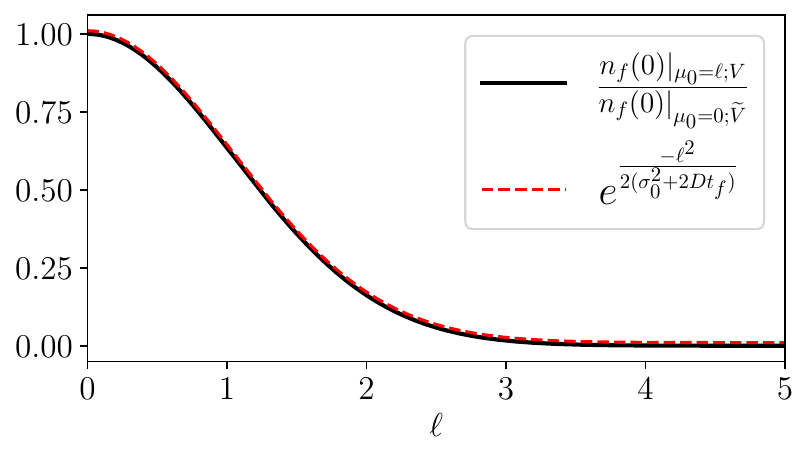}
    \caption{The ratio of $n_f(0)|_{\mu_0=\ell;V}$, obtained from numerically solving 
    Eq.~\eqref{eq:mean-field} in $d=1$ dimensions for three fitnesses $V(x,t)$ (see below) starting from the germline~\eqref{eq:n0-def} 
    $n_f(0)|_{\mu_0=0;\widetilde{V}}$, to that evaluated from Eq.~\eqref{eq:mean-field} with the modified fitness $\widetilde{V}$ (see Eq.~\eqref{eq:V-tilde-def}) and starting from the modified initial population $N_0 \mathcal{G}(x_0,\sigma_0^2)$  instead centered on the target.  
    The solid curve is obtained from the numerics for a set of arbitrary fitness functions ($V_{\mathrm{I}} = \mathrm{max}(1-|x|,0)$,  
    $V_{\mathrm{II}} = \mathrm{min}((x-2+t)^2,1)$,  
    and $V_{\mathrm{III}}= \mathcal{G}(x- x^*,\sqrt{3 D (t_f - t)})$) and their respective $\widetilde{V}$.}
    \label{fig:path-integral-ratio}
\end{figure}

Noting that
$\dot{\tilde{\bm{r}}}(t) = \dot{\bm{r}}(t) + \bm{v}$, the path integral action (Eq.~\eqref{eq:path-integral-action}) transforms as follows
\begin{align}   \label{eq:action-transformed}
    S[\bm{r},\dot{\bm{r}}] &= \frac{\bm{v}^2 t_f}{4D} - \frac{\bm{v}\cdot \left(\tilde{\bm{r}}(t_f)-\tilde{\bm{r}}(0) \right)}{2D} 
    \nonumber \\
    &\qquad +  \int_0^{t_f} du \left(\frac{\dot{\tilde{\bm{r}}}^2(u)}{4D} - \widetilde{V}\left(\tilde{\bm{r}}(u),u\right) \right) \, 
    \\
    &\equiv \frac{\bm{v}^2 t_f}{4D} - \frac{\bm{v}\cdot\left(\tilde{\bm{r}}(t_f)-\tilde{\bm{r}}(0) \right)}{2D}  +
    \widetilde{S}[\tilde{\bm{r}},\dot{\tilde{\bm{r}}}]
\end{align}
where we have used the fact that 
$\int_0^{t_f} du \, \dot{\tilde{\bm{r}}}(u) = \tilde{\bm{r}}(t_f) - \tilde{\bm{r}}_0$. 
Here, 
$\widetilde{V}\left(\tilde{\bm{r}}(u),u\right)$ and $\widetilde{S}[\tilde{\bm{r}},\dot{\tilde{\bm{r}}}]$ denote the transformed vaccine fitness and path integral action which are obtained from $V\left(\bm{r}(u),u\right)$ and $S[\bm{r},\dot{\bm{r}}]$ by making the change 
$\bm{r}(u) \to \tilde{\bm{r}}(u)+\bm{z}(u)$. 
Substituting these expressions back into Eq.~\eqref{eq:path-integral-solution}, we obtain for $\bm{x}=0$ 
\begin{align}   \label{eq:path-integral-midstep}
    &n_f(0) = 
    \exp{ \frac{\bm{v}^2 t_f}{4D} - \frac{\bm{v}\cdot\bm{z}(t_f)}{2D} }
    \nonumber \\
    &\times e^{-\lambda t_f} \int d^d\tilde{\bm{x}}_0 \, 
    \tilde{n}_0 (\tilde{\bm{x}}_0) \, e^{ \frac{\bm{v}\cdot\tilde{\bm{x}}_0}{2D}} \, 
    \int_{\tilde{\bm{r}}(0)=\tilde{\bm{x}}_0}^{\tilde{\bm{r}}(t_f) = \bm{z}(t_f) } 
    \mathcal{D}\tilde{\bm{r}}(u) \, 
    e^{-\widetilde{S}[\tilde{\bm{r}},\dot{\tilde{\bm{r}}}]} \, .
\end{align}
To proceed, we utilize the Gaussian shape of $\tilde{n}_0$ and simplify the outer integral by combining the modified initial population $\tilde{n}_0(\tilde{\bm{x}}_0) = N_0\mathcal{G}(\tilde{\bm{x}}_0-\tilde{\bm{\mu}}_0, \sigma_0^2)$ with the exponential factor, $e^{ \frac{\bm{v}\cdot \tilde{\bm{x}}_0}{2D}}$, 
through completing the square, which yields
\begin{align}   \label{eq:n0-shift}
    \tilde{n}_0(\tilde{\bm{x}}_0) \, e^{\frac{\bm{v}\cdot\tilde{\bm{x}}_0}{2D}} =   N_0\mathcal{G}\left(\tilde{\bm{x}}_0 - \tilde{\bm{\mu}}_0 + \frac{\bm{v}\sigma_0^2}{2D} , \sigma_0^2 \right) \, 
    e^{-\frac{\bm{v}\cdot\tilde{\bm{\mu}}_0}{2D} + \frac{\bm{v}^2 \sigma_0^2}{8D^2}} \, . 
\end{align}
We observe that apart from the multiplicative constant,  the transformed path integral expression in Eq.~\eqref{eq:path-integral-midstep} retains the same mathematical structure as the original expression in  Eq.~\eqref{eq:path-integral-solution}.

It is now straightforward to show that for $\bm{z}_0 = \bm{v} t_f = \frac{\bm{\mu}_0}{1+\frac{\sigma_0^2}{2Dt_f}}$, we have $\bm{z}(t_f)=0$ and, moreover,  $\tilde{\bm{\mu}}_0 - \frac{\bm{v}\sigma_0^2}{2D} = 0$. Substituting these back into Eq.~\eqref{eq:path-integral-midstep} and using Eq.~\eqref{eq:n0-shift}, we finally arrive at
\begin{align} \label{eq:path-integral-final-full}
    &n_f(0) = \exp{-\frac{\bm{\mu}_0^2}{2(\sigma_0^2+2Dt_f)}} 
    \nonumber \\
    &\quad \times 
    N_0 e^{-\lambda t_f}
    \int d^d\bm{x}_0 \, \mathcal{G}(\bm{x}_0,\sigma_0^2) 
    \int_{\bm{r}(0)=\bm{x}_0}^{\bm{r}(t_f)=0} 
    \mathcal{D}\bm{r}(u) \, 
    e^{-\widetilde{S}[\bm{r},\dot{\bm{r}}]} \, , 
\end{align}
where we have reverted the dummy integration variables $\tilde{\bm{x}}_0$ and $\tilde{\bm{r}}$ back to $\bm{x}_0$ and $\bm{r}$. 
In Eq.~\eqref{eq:path-integral-final-full}, the second line on the right-hand side can be recognized as the right-hand side of the original expression in Eq.~\eqref{eq:path-integral-solution} upon changing $V \to \widetilde{V}$ and setting $\bm{\mu}_0 =0$. Rearranging Eq.~\eqref{eq:path-integral-final-full} thus leads to the Eq.~\eqref{eq:nfx0-ratio-formula} in the main text. 
Figure~\ref{fig:path-integral-ratio} shows that the left-hand side of Eq.~\eqref{eq:nfx0-ratio-formula}, evaluated by numerically solving the one-dimensional mean-field equation for a set of arbitrary vaccine fitness functions, matches the exponential factor on the right-hand side of Eq.~\eqref{eq:nfx0-ratio-formula}.

%% file: Appendix_BCH-discrete.tex
Since the SSA are performed on a system of discrete bins, for completeness, this appendix demonstrates how the calculations 
in Section~\ref{sec:bch} can be extended to the discrete-bin version of the mean-field dynamics. 

Recall that the general mean-field dynamics for the discretized bins before taking the continuum limit is given by Eq.~\eqref{eq:mean-field-discrete}. We set $\lambda_{\bm{x}} = \lambda$ and $\gamma_{\bm{x}\bm{y}} = \gamma \sum_{k=1}^d \delta_{\bm{x}, \bm{y}\pm \hat{e}_k}$ where $\hat{e}_k$ is the $k$th unit vector and $d$ denotes the dimension of  shape space. 
Focusing on the one-dimensional case for simplicity, we use the BCH formula~\eqref{eq:BCH-def} with $A$ and $B$ matrices (instead of continuous operators) chosen as $A \to \widetilde{V}_{\ell \ell'} (t) = \widetilde{V}_\ell (t) \delta_{\ell, \ell'}$ for 
$\ell \in \{0,1,2,\ldots,N_{\mathrm{bin}}-1\}$, representing the discrete fitness function,  
and $B  \to D_{\ell\ell'} = \gamma (\delta_{\ell,\ell'+1} - 2 \delta_{\ell,\ell'} + \delta_{\ell,\ell'+1})$ as the discrete mutation matrix (note that the diagonal terms at $\ell =1$ and $\ell=N_{\mathrm{bin}}$ in the mutation matrix should be set to $-\gamma$). 
It is then straightforward to obtain the corresponding commutator matrix as 
$C_{\ell\ell'} = \gamma (\widetilde{V}_{\ell'} - \widetilde{V}_\ell )(\delta_{\ell,\ell'+1}+\delta_{\ell,\ell'-1})$. Following through the same steps leading to Eq.~\eqref{eq:BCH_nfinal}, one can obtain the BCH approximation for the bnAb count in the discrete-bin model as 
\begin{align} 
    n_{\ell=0}(t_f)  \approx \quad & 
    N_0 e^{-\lambda t_f} 
    e^{\int_0^{t_f}dt \widetilde{V}_{\ell=0}(t)}  \nonumber \\
    \times 
    &e^{\gamma \int_0^{t_f} dt (t_f-t) \big(\widetilde{V}_{\ell=1}(t) - 2\widetilde{V}_{\ell=0}(t) + \widetilde{V}_{\ell=-1}(t) \big)  } \nonumber \\
    \times &[e^{t_f D_{\ell'\ell''}} n_{\ell''}(t=0)]. 
\end{align}
For the Gaussian fitness profile, optimizing this expression leads to the following transcendental equation
\begin{align}   \label{eq:BCH-discrete-transcendental}
    \frac{ \exp{ \frac{1}{2\sigma_V^2(t)}}}{\frac{1}{\sigma_V^2(t)}-1} = \frac{2\gamma(t_f-t)}{1-2\gamma(t_f-t)} . 
\end{align}
This equation has a unique solution for 
$2\gamma(t_f-t) \geq \frac{1}{1+2e^{-3/2}}$ and no solution otherwise, defining a temporal window preceding the final time. It can be readily verified that the optimal width that solves Eq.~\eqref{eq:BCH-discrete-transcendental} decreases  over time until it reaches this final temporal window, beyond which the width remains fixed. Therefore, within the discrete-bin dynamics, the optimal fitness spread automatically reaches a finite minimum value, preventing the fitness amplitude from becoming singular, as discussed in Section~\ref{sec:bch}.

%% file: Appendix_trajectory.tex
 \section{Time evolution of B cell population under different vaccine protocols} 
 \label{appendix:trajectory}

\begin{figure}
 \begin{minipage}[c]{.85\linewidth}
		\centering
		\includegraphics[width=\linewidth]{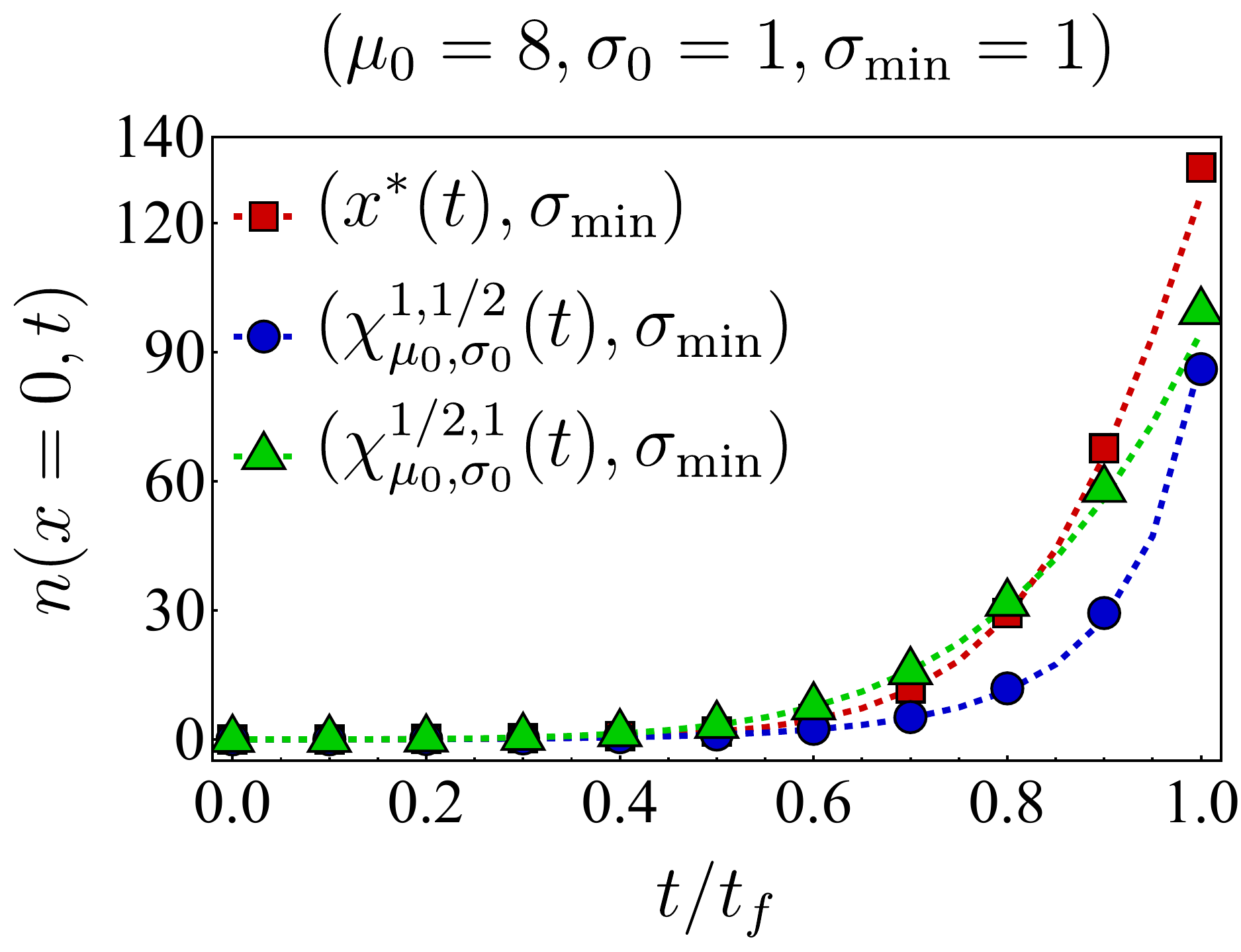}
	\end{minipage} 
 \begin{minipage}[c]{.87\linewidth}
		\centering
		\includegraphics[width=\linewidth]{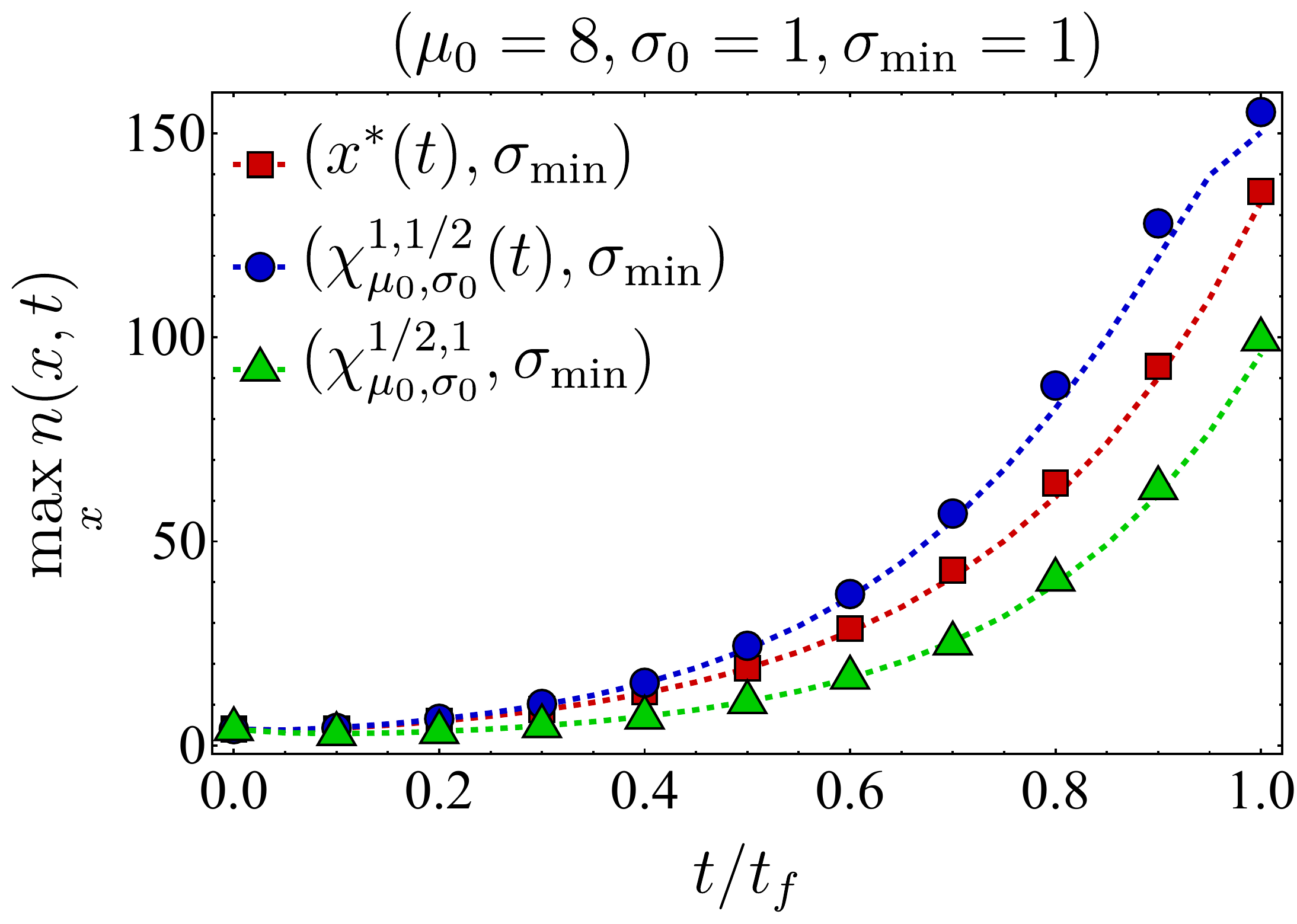}
	\end{minipage}  
\caption{
Plots of bnAb count (top) and the peak size of the B cell population (bottom) as functions of time, utilizing the same parameters as in Fig.~\ref{fig:nf0-vs-mu0}. The dashed lines represent the numerical solutions to the continuum mean-field equation~\eqref{eq:mean-field} whereas the solid points depict averages over stochastic realizations of the discrete master equation~\eqref{eq:master-fock} sampled using SSA. }
   \label{fig:n-vs-time}
\end{figure}

\begin{figure*}[t]
 \begin{minipage}[c]{.45\linewidth}
		\centering
		\includegraphics[width=\linewidth]{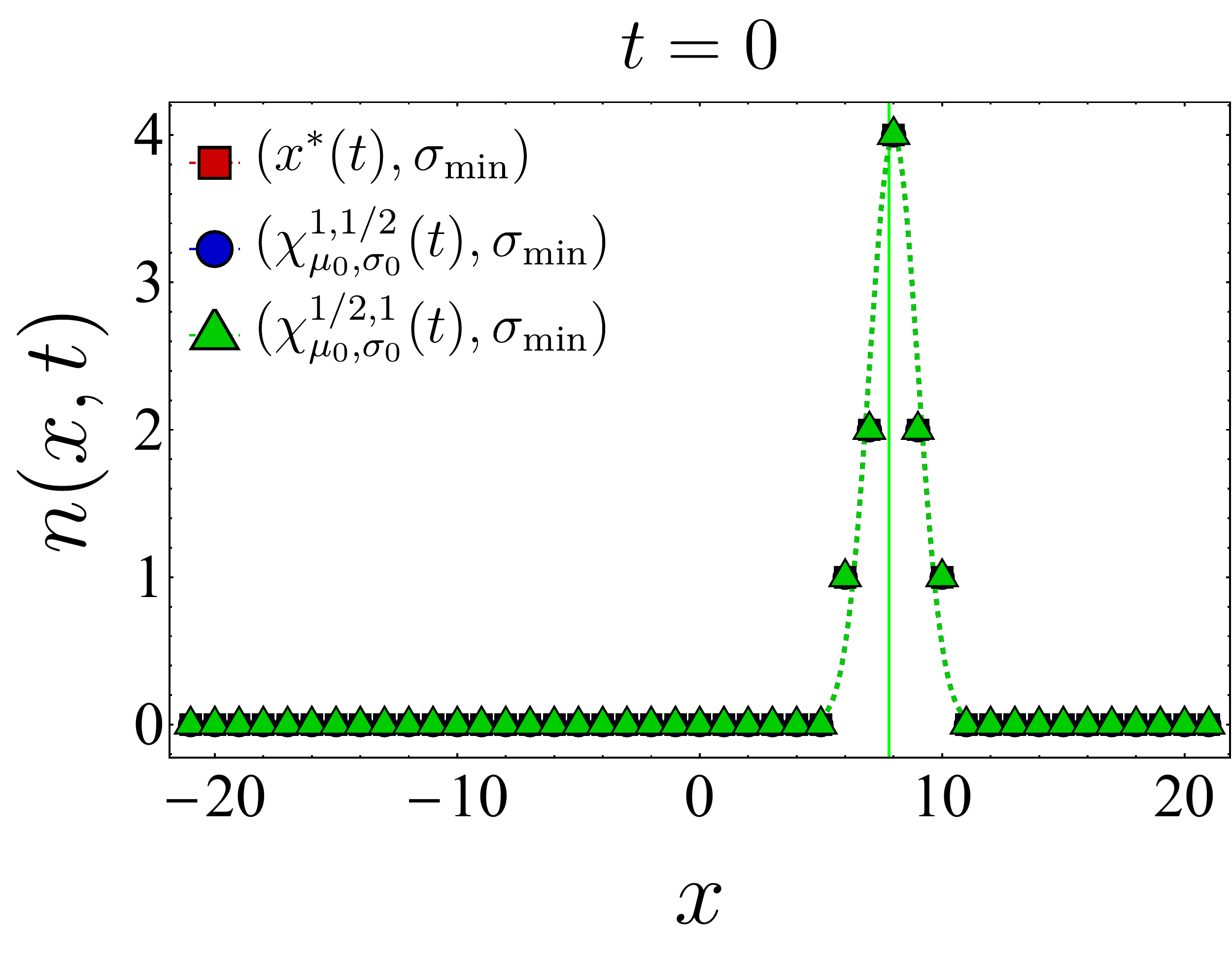}
	\end{minipage} 
	\hskip.25cm
 \begin{minipage}[c]{.45\linewidth}
		\centering
		\includegraphics[width=\linewidth]{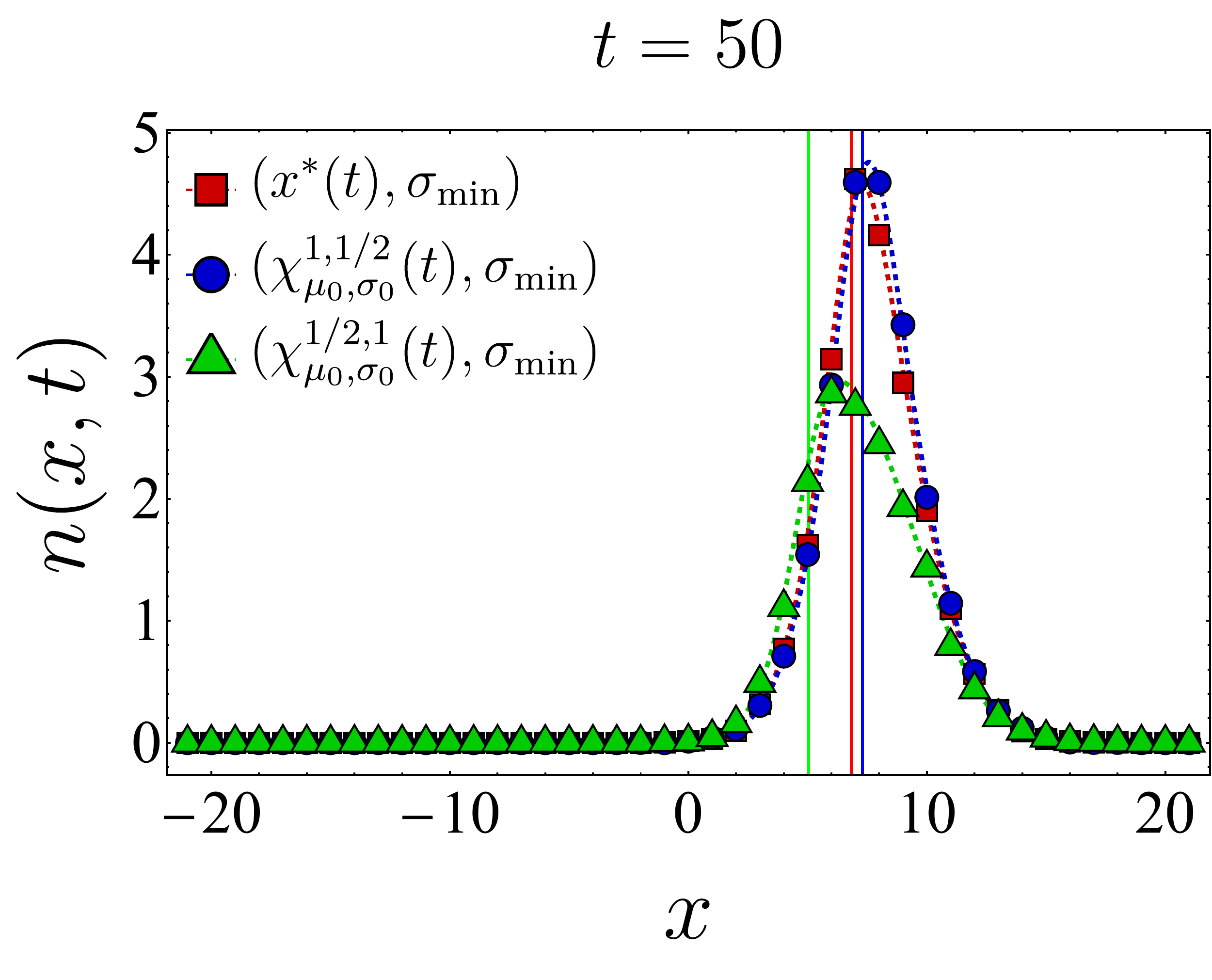}
	\end{minipage} 
 	\hskip.25cm
   \begin{minipage}[c]{.45\linewidth}
		\centering
		\includegraphics[width=\linewidth]{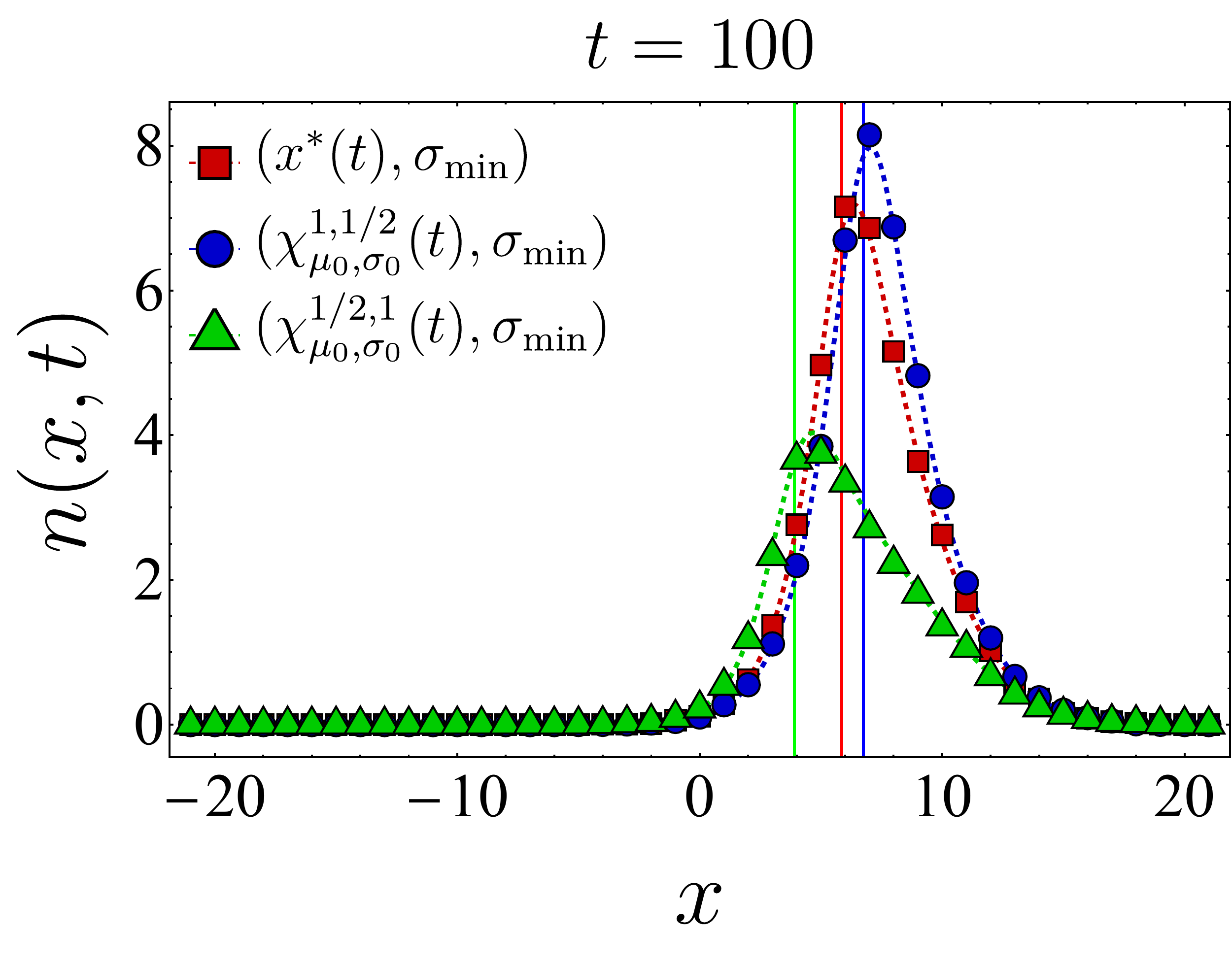}
	\end{minipage} 
 \begin{minipage}[c]{.45\linewidth}
		\centering
		\includegraphics[width=\linewidth]{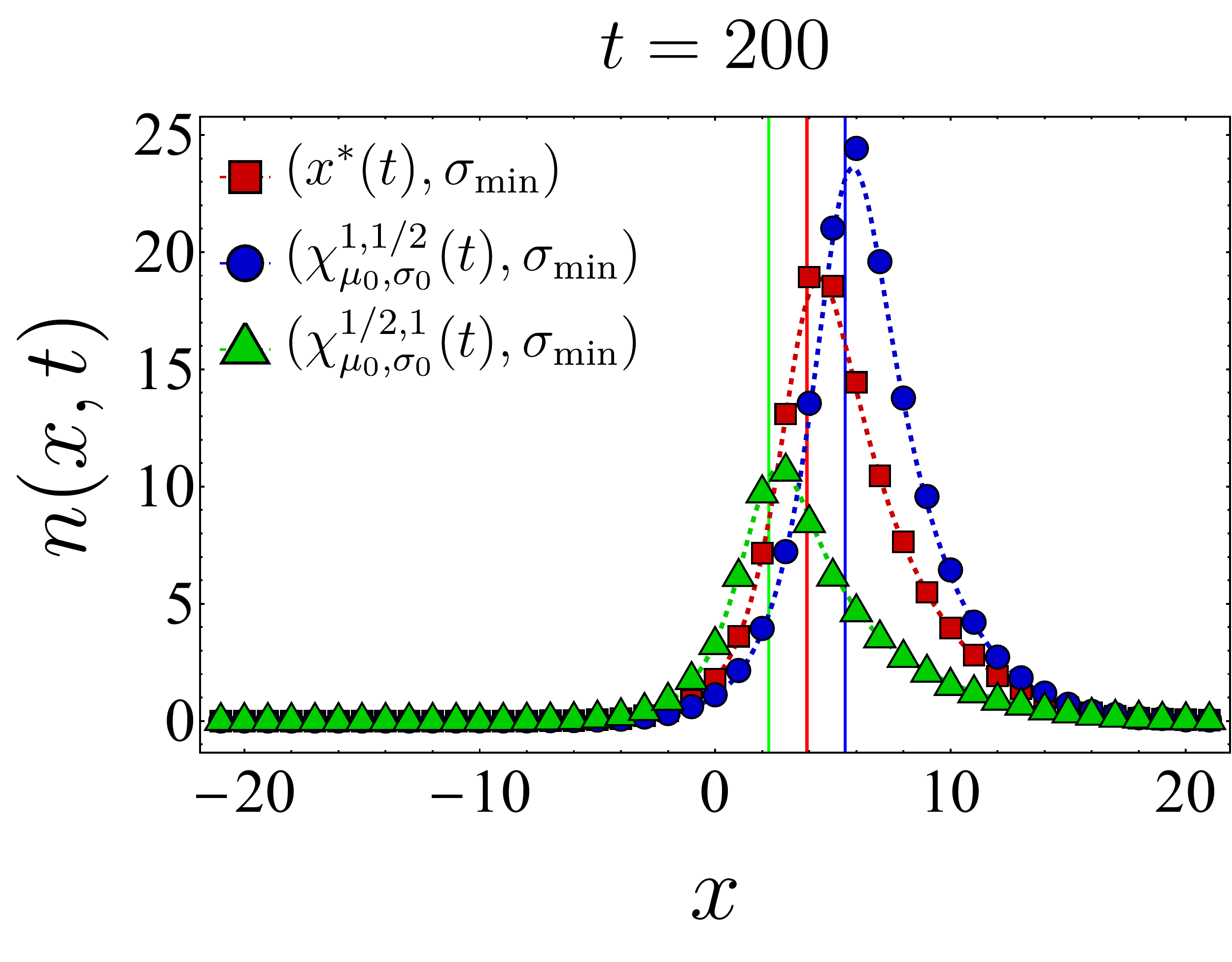}
	\end{minipage} 
	\hskip.25cm
 \begin{minipage}[c]{.45\linewidth}
		\centering
		\includegraphics[width=\linewidth]{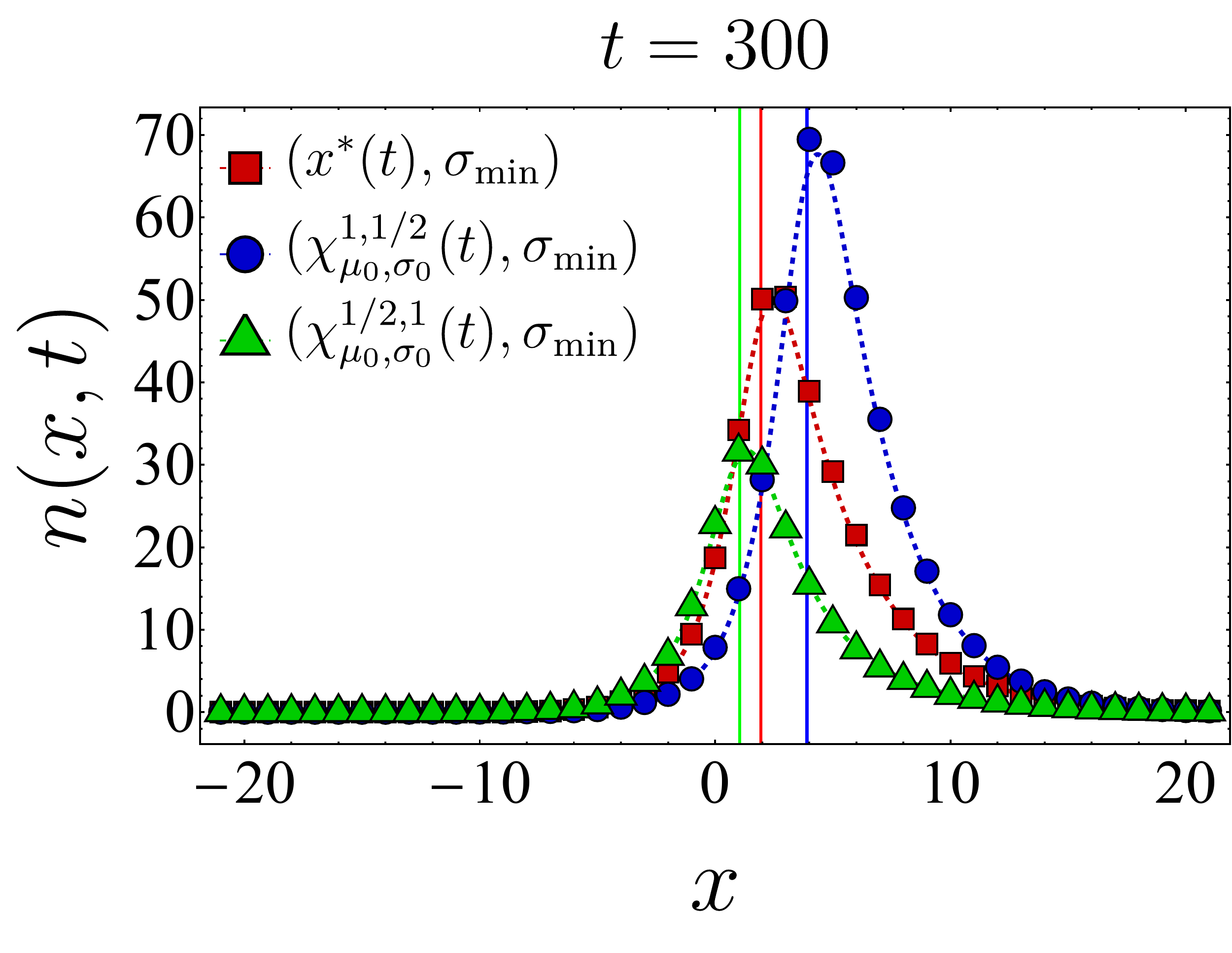}
	\end{minipage} 
 	\hskip.25cm
   \begin{minipage}[c]{.45\linewidth}
		\centering
		\includegraphics[width=\linewidth]{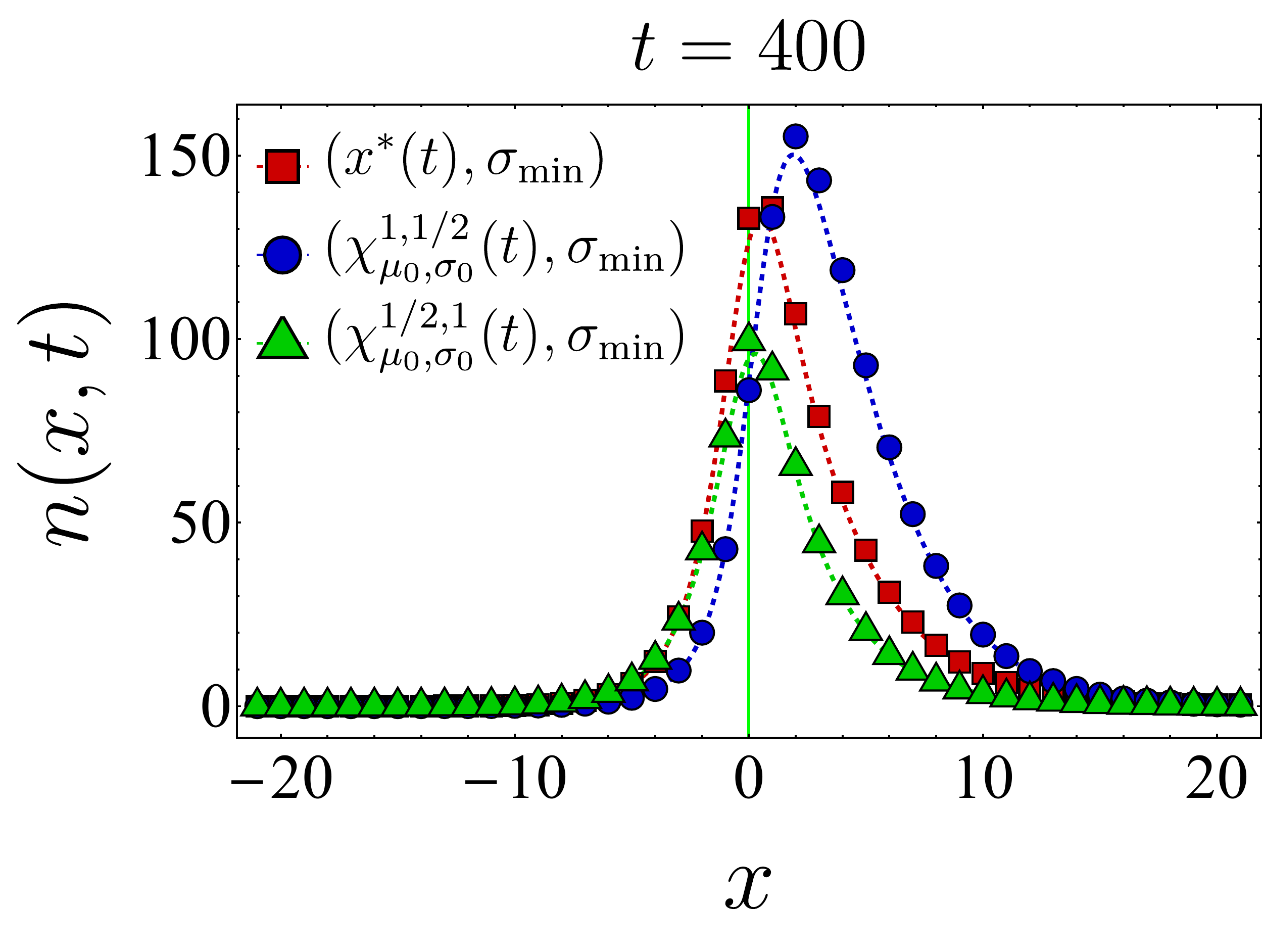}
	\end{minipage} 
\caption{
Snapshots of the B cell population at different times, corresponding to the vaccination protocols of the left panel in Fig.~\ref{fig:nf0-vs-mu0}. In each case, the dashed curves represent the mean-field numerical solution, while the solid data points depict the SSA sample averages taken over 1000 realizations. The vertical lines in different colors (red, blue, green) indicate the location of the (moving) vaccine center for the corresponding protocol. }
   \label{fig:trajectories}
\end{figure*}

Here, we present further computational results on the time evolution of the B cell population in the one-dimensional shape space. 
Figure~\ref{fig:n-vs-time} shows the bnAb count (top) and population peak's size (bottom) as functions of time. These results correspond to the same vaccination protocols and parameters as the left panel of Fig.~\ref{fig:nf0-vs-mu0}. The graphs suggest that although positioning the vaccine-induced fitness peak optimally leads to the highest final bnAb count, the resulting population peak remains smaller compared to that achieved with the  $\chi^{1,1/2}_{\mu_0,\sigma_0}(t)$ protocol. 
This observation is reinforced by the population snapshots in Fig.~\ref{fig:trajectories}, indicating that while the $x^*(t)$  protocol effectively maximizes $n_f(0)$, it may not fully optimize other aspects of the final B cell population, such as its total size or peak.

\begin{figure}
    \centering
    \includegraphics[width=.84\linewidth]{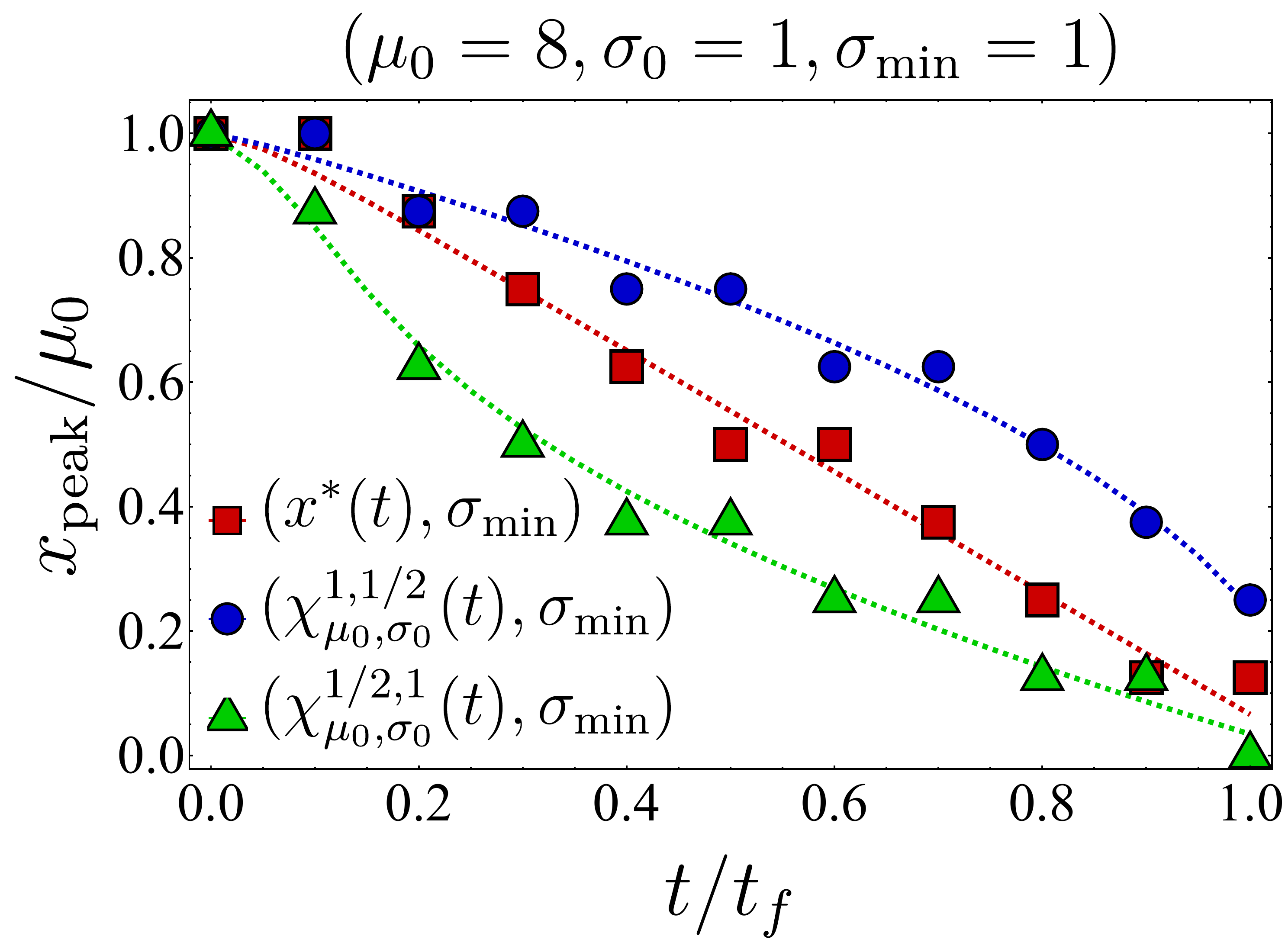}
    \caption{The location of the population peak in the one-dimensional shape space as a function of time and for three different choices of the vaccine center. Note that in the stochastic simulations, the population peak can only move between the discrete bins, resulting in step-like jumps and periods of no motion in the corresponding data (solid points). }
    \label{fig:xPeak-vs-t}
\end{figure}

In Fig.~\ref{fig:xPeak-vs-t}, we present the location of the population's peak as a function of time for the same vaccination protocols. Notably, only for the $(\chi^{1/2,1}_{\mu_0,\sigma_0},\sigma_{\min})$ protocol, the population peak appears to reach the target bin, whereas for the other two protocols, the peak remains away from $x=0$ at the final time.   
For the optimal vaccine center, and after an initial transient period, the population peak consistently trails $x^*(t)$ with a constant lag. Conversely, with other choices, the population's distance from the vaccine center fluctuates over time. As discussed following Eq.~\eqref{eq:xv-optimal-xstar}, such fluctuations in distance may either result in a slow movement of the population towards the target when the vaccine center remains too close to the population peak (blue) or lead to limited population growth when the vaccine center moves too far from the population (green).

